\documentclass[nohyper,12pt,letterpaper]{JHEP3}

\date{January, 2003}

\title{Twisted Backgrounds, PP-Waves and Nonlocal Field Theories}

\author{
{Mohsen Alishahiha$^a$ and Ori J. Ganor$^b$}
\vspace*{5mm}

{\it $^a\ $ Institute for Studies in Theoretical Physics and
Mathematics (IPM)} \\
{\it P.O.Box 19395-5531, Tehran, Iran}\\
E-mail: \email{alishah@theory.ipm.ac.ir}\\

\vspace*{2mm}

{\it $^b\ $ Department of Physics, 366 Le Conte Hall}\\
{\it University of California, Berkeley, CA 94720}\\

\vspace*{2mm}

{\it and}\\

\vspace*{2mm}

{\it Lawrence Berkeley National Laboratory}\\
{\it Berkeley, CA 94720}\\
E-mail: \email{origa@socrates.berkeley.edu}

} 

\abstract{
We study partially supersymmetric plane-wave like deformations
of string theories and M-theory on brane backgrounds.
These deformations are dual to nonlocal field theories. 
We calculate various expectation values of configurations 
of closed as well as open Wilson loops and Wilson surfaces
in those theories. We also discuss the manifestation of the
nonlocality structure in the supergravity backgrounds.
A plane-wave like deformation of little string theory 
has also been studied.
} 


\received{January 29, 2003}

\preprint{IPM/P-2003/002 \\
          UCB-PTH-02/60  \\
          LBL-51922      \\
          \hepth{0301080} \\
}  
\begin{document}
\def\be{\begin{equation}} 
\def\ee{\end{equation}} 
\def\bear{\begin{eqnarray}} 
\def\eear{\end{eqnarray}} 
\def\nn{\nonumber} 

\def\defineas{\stackrel{\mbox{\tiny \rm def}}{=}}
\def\hlf{{{1\over 2}}} 
 
\def\eqdf{{\stackrel{def}{=}}} 
\def\const{{\mbox{const\ }}}                     
 
\def\lbr{{\lbrack}} 
\def\rbr{{\rbrack}} 
\def\dg{{\dagger}} 
\def\wdg{{\wedge}} 
 
\def\a{\alpha} 
\def\b{\beta} 
\def\g{\gamma} 
\def\u{\mu} 
\def\v{\nu} 
\def\s{\sigma}
\def\t{\tau}
\def\r{\rho} 
\def\th{{\theta}} 
\def\lam{{\lambda}} 

\def\bth{{\overline{\theta}}} 
\def\blam{{\overline{\lambda}}} 
\def\bpsi{{\overline{\psi}}} 
\def\wpsi{{\widetilde{\psi}}} 
\def\bwpsi{{\overline{\widetilde{\psi}}}}

\def\bsig{{\overline{\sigma}}} 

\def\ad{{\dot{\alpha}}} 
\def\bd{{\dot{\beta}}} 
\def\gd{{\dot{\gamma}}} 
 
\def\Lbd{{\cal L}} 
\def\bj{{\overline{j}}} 
\def\bk{{\overline{k}}} 
\def\bz{{\overline{z}}} 
\def\wF{{\widetilde{F}}} 
\def\wA{{\widetilde{A}}}

\def\rt{{\rightarrow}}  

\def\cc{{\mbox{c.c.}}} 

\newcommand{\rep}[1]{{{\bf {#1}}}}      
\newcommand{\tr}[1]{{\mbox{tr}\{{#1}\}}}          
\newcommand{\ttr}[1]{{\mbox{Tr}\{{#1}\}}}          
\newcommand{\trr}[2]{{\mbox{tr}_{#1}\{{#2}\}}}    
\newcommand{\ev}[1]{{\langle {#1} \rangle}}             
\newcommand{\evtr}[1]{{\lgl \tr{{#1}} \rgl}}      
\newcommand{\com}[2]{{\lbrack {#1},{#2}\rbrack}}  
\newcommand{\acom}[2]{{\{ {#1},{#2} \}}}          
\newcommand{\cov}[1]{{\nabla_{#1}}}               
\newcommand{\bra}[1]{{\langle {#1}|}}
\newcommand{\ket}[1]{{|{#1}\rangle}}
\newcommand{\inner}[2]{{\langle {#1} | {#2}\rangle}}

\newcommand\px[1]{{\partial_{#1}}} 
\newcommand\qx[1]{{\partial^{#1}}} 
\newcommand\QD[1]{{{\cal Q}{\mbox{D}}_{\lbrack{#1}\rbrack}}}
\newcommand\QKK[1]{{{\cal Q}{\mbox{KK}}_{\lbrack{#1}\rbrack}}}
\newcommand\QFS[1]{{{\cal Q}{\mbox{FS}}_{\lbrack{#1}\rbrack}}}
\newcommand\QNS[1]{{{\cal Q}{\mbox{NS5}}_{\lbrack{#1}\rbrack}}}
\newcommand\QM[2]{{{\cal Q}{\mbox{M{#1}}}_{\lbrack{#2}\rbrack}}}

\newcommand\VD[1]{{{\cal V}{\mbox{D}}_{\lbrack{#1}\rbrack}}}
\newcommand\VFS[1]{{{\cal V}{\mbox{FS}}_{\lbrack{#1}\rbrack}}}
\newcommand\VNS[1]{{{\cal V}{\mbox{NS5}}_{\lbrack{#1}\rbrack}}}
\newcommand\VM[2]{{{\cal V}{\mbox{M{#1}}}_{\lbrack{#2}\rbrack}}}

\newcommand\QJ[1]{{J_{\lbrack{#1}\rbrack}}}

\newcommand{\SUSY}[1]{{{\cal N}= {#1}}}           
\newcommand{\SLZ}[1]{{SL({#1},\Z)}}              
\newcommand{\SLR}[1]{{SL({#1},\R)}}              
\newcommand{\Spin}[1]{{{\mbox{Spin}}({#1})}}      

\def\gYM{g_{\mbox{\tiny \rm YM}}} 
\def\Leff{{L_{\mbox{\tiny \rm eff}}}} 
\def\mueff{{\mu_{\mbox{\tiny \rm eff}}}} 

\def\spX{{\cal X}}           

\def\mathbb{\bf}
\newcommand{\Z}{{\mathbb Z}}
\newcommand{\R}{{\mathbb R}}
\newcommand{\C}{{\mathbb C}}
\newcommand\MR[1]{{\R^{#1}}}
\newcommand\MC[1]{{\C^{#1}}}

\newcommand\SetComp[1]{\left\{\!\!\left\{{#1}\right\}\!\!\right\}}

\def\Om{{\Omega}}

\def\wstar{{\widetilde{\star}}}
\def\bk{{\overline{k}}}

\def\mcr{\mathcal{R}}
\def\xv{\vec{x}}
\def\xvt{\vec{x}^{\top}}
\def\nn{\nonumber}
\def\nv{\hat{n}}
\def\nvt{\hat{n}^{\top}}
\def\Ot{\Omega^{\top}}

\def\valQ{{{\cal U}}}                

\def\cO{{{\cal O}}}           
\def\cwO{\widetilde{{\cal O}}}   
\def\wF{{\widetilde{F}}}       
\def\hPhi{{\hat{\Phi}}}   
\def\wPhi{{\widetilde{\Phi}}}   
\def\bPhi{{\overline{\Phi}}}   
\def\hbPhi{{\hat{\overline{\Phi}}}}   
\def\Path{{{\cal P}}}   
\def\tw{{\eta}}         
\def\bpsi{{\overline{\psi}}}     
\def\wpsi{{\widetilde{\psi}}}
\def\Lag{{\cal L}}       
\def\wLag{\widetilde{{\cal L}}}       

\def\vL{{\vec{L}}}
\def\bZ{{\overline{Z}}}

\def\Re{{\rm Re\hskip0.1em}}
\def\Im{{\rm Im\hskip0.1em}}

\def\Id{{\bf I}}                                

\def\cpl{{\lambda}}  
\def\mcr{{\mathcal{R}}}
\def\xv{{\vec{x}}}
\def\xvt{{\vec{x}^{\top}}}
\def\nv{{\hat{n}}}
\def\nvt{{\hat{n}^{\top}}}
\def\hM{{M}}
\def\hMt{{\hM^\top}}
\def\bM{{\bar{M}}}
\def\bMt{{\bM^\top}}
\def\cM{{\cal M}}
\def\cMt{{\cM^\top}}
\def\tM{{\widetilde{M}}}
\def\utr{{\mbox{tr}}}

\def\mQ{{{m}}}          
\def\fb{{\varphi}}      
\def\tfb{{\tilde{\varphi}}}      
\def\Wfb{{\Phi}}       
\def\yb{{y}}      
\def\Wyb{{Y}}       
\def\bgs{{{\bar g}_s}}      
\def\hDen{{\Delta}} 

\def \PP {{\cal P}} 

\def\xpr{\mathaccent 19 x}  
\def\Ypr{\mathaccent 19 Y}  
\def\etpr{\mathaccent 19 \eta}  
\def\thpr{\mathaccent 19 \theta}  

\def\wxmin{{\tilde{x}^{-}}} 
\def\Rchg{{{\cal Q}}}    

\section{Introduction}\label{sec:intro}

The pp-wave metric 
$$
ds^2 = dx^{+} dx^{-} -\sum_{i=1}^{D-2} (dx^i)^2 
-K(x_1, \dots, x_{D-2}) (dx^{+})^2
$$
is a deformation of the $(D-1)+1$ dimensional Minkowski metric with the  
property
that all the scalars that can be constructed out of the
curvature tensor are identically zero.
Realizations of such metrics in string theory have been extensively studied 
recently (see for example \cite{Melvin:1963qx}-\cite{Berenstein:2002jq}).
In string theory such a metric can be realized if there is also
a nonzero RR or NSNS flux.
For example, the following equation describes a good type-II background 
when the string coupling constant $g_s = 0$,
\bear
ds^2 &=& dx^{+} dx^{-} -d\xvt d\xv 
-{\a'}^{-2}(\xvt\cMt\cM\xv)(dx^{+})^2,\qquad
H = \eta^{-1} {\a'}^{-2}d\xvt\wdg\cM d\xv\wdg dx^{+},\nn\\
&& \xv \defineas (x^1,\dots,x^8).
\label{Hppwave}
\eear
Here $H$ is a 3-form NSNS flux (for $\eta = 1$) or RR flux
(for $\eta = g_s$, the string coupling constant)
and $\cM$ is a constant antisymmetric $8\times 8$ matrix.
Such backgrounds have been constructed in
\cite{{Figueroa-O'Farrill:2001nz},{Maldacena:2002fy},{Russo:2002qj}}.
In lightcone gauge the worldsheet theories that correspond to such
backgrounds are free.
To see this in the case of NSNS flux requires a change of variables 
\cite{Russo:2002rq} ( see also \cite{Michishita:2002jp}),
and the case of RR flux was demonstrated in \cite{Metsaev:2001bj}.

One can also find a deformation of the $AdS_5\times S^5$
metric by a $(dx^+)^2$ term as follows:
\bear
ds^2 &=& \frac{R^2}{r^2}(dx^{+} dx^{-} -d\xvt d\xv -dr^2)
-\frac{R^2}{r^4}(\nvt\cMt\cM\nv)(dx^{+})^2 - R^2 d\nvt d\nv,\nn\\
B &=& \frac{1}{r^2}d\nvt\cM\nv\wdg dx^{+}.
\label{HppAdS}
\eear
Here $\nv$ is a unit vector in $\R^6$ that parameterizes the $S^5$ of
radius $R$ and
we have written the NSNS 2-form field potential $B$ instead of the
field strength.
There is also a 5-form flux. It is also possible to replace the NSNS
2-form $B$ in (\ref{HppAdS}) with an RR 2-form field with exactly the
same form but with a prefactor of $g_s^{-1}$.
Backgrounds similar to those have been recently discussed in
\cite{Hubeny:2002nq}. This background has the property that all the
scalars that can be constructed out of the curvature tensor are identical
to those of the undeformed $AdS_5\times S^5$ space.
Since $AdS_5\times S^5$ is dual to $\SUSY{4}$ $SU(N)$ Super-Yang-Mills theory
\cite{Maldacena:1997re} it is interesting to find out what field theory
is dual to the supergravity background (\ref{HppAdS}).\footnote{
Background similar to (\ref{HppAdS}) has also been studied in
\cite{Alishahiha:2000pu} in the context of lightlike noncommutativity
field theory \cite{Aharony:2000gz}.} It would be also
interesting to generalize
this deformation for other brane backgrounds \cite{Itzhaki:1998dd}.

One of the purposes of this paper is to describe the field theory duals of
(\ref{HppAdS}) and other backgrounds that can be similarly described as a
deformation
of $AdS_q\times S^p$ with a $(dx^{+})^2$ in the metric and with lightlike 
[i.e. of the form $(\cdots)\wdg dx^+$] fluxes.
We will find that (\ref{HppAdS}) describes the large $N$ limit of a gauge
theory with nonlocal interactions. We will describe how the nonlocal
interactions
manifest themselves in (\ref{HppAdS}) and we will study expectation values of
various configurations of Wilson loops in the theory.

Another purpose of this paper is to observe that nonlocal deformations of
field theories are a rather general phenomenon that occurs when
$AdS_q\times S^p$ is deformed with extra Lorentz invariance breaking fluxes.
We will describe a deformation of $AdS_7\times S^4$ that corresponds
to a nonlocal deformation of the $(2,0)$ theory and the nonlocality is
described by a parameter that is a 2-form.
We will also discuss various deformations of the 5+1D little string theory 
\cite{{Berkooz:1997cq},{Seiberg:1997zk}}.

The paper is organized as follows.
In sections \ref{sec:twback}-\ref{sec:sugra}
we present a set of backgrounds that can be obtained
from flat space by a simple construction. These ``twisted'' backgrounds will
serve as the ambient space for brane probes in section \ref{sec:probes}.
We will argue that the theories on the probes are in general nonlocal.
We will study their large $N$ limit in section \ref{sec:LargeN}.
There we will discover that the backgrounds such as (\ref{HppAdS}) are
the supergravity duals of nonlocal field theories.
We follow in section \ref{sec:Wilson} with a study of Wilson loops
and their generalizations in such backgrounds.
In section \ref{sec:nonlocality} we will demonstrate various manifestations
of the nonlocal nature of the supergravity backgrounds. The last section
is devoted to the discussion.

\section{Construction of twisted backgrounds}\label{sec:twback}
Our goal in this section is to construct backgrounds for M-theory
and string theory.
These will be non-geometrical backgrounds where brane probes are,
in general, described by nonlocal field theories.
These backgrounds are a subset of those constructed in
\cite{Figueroa-O'Farrill:2001nz,Gutperle:2001mb,Figueroa-O'Farrill:2001nx}
and we will review them below.
To construct the backgrounds we 
start with M-theory or string-theory compactified on $T^d$.
We then continuously deform the geometry while
keeping the space locally flat. 
We do this by introducing a  ``geometrical'' twist.
Using T-duality or U-duality on these backgrounds
we will then construct non-geometrical twists.

\subsection{Geometrical twists}
We will use the term ``geometrical twist''  to refer to a 
compactification on $T^d$ with nontrivial Wilson loops
for the transverse $\Spin{9-d}$ (in string theory, or
in M-theory $\Spin{10-d}$) rotation group.
Take $d$ commuting elements
$$
\Omega_1,\Omega_2,\dots,\Omega_d\in\Spin{9-d}
$$
and define the space that is locally $\R^{9,1}$
but with the global identifications
\bear
(x_0,\,x_1,\dots, x_d,\,\xv) &\sim&
(x_0,\,
x_1 + 2\pi n_1 R_1,
\dots,
x_d + 2\pi n_d R_d,\,
\prod_{j=1}^d\Omega_j^{n_j}\xv),
\nn\\ &&
\xv \defineas (x_{d+1},\dots,x_9).
\label{tdtw}
\eear
A special case of this background is $\MR{9,1}$ with the identification
\be\label{sotw}
(x_0, x_1,\, \xv) \sim 
(x_0,x_1 + 2\pi R,\,\Omega\xv)
\quad
\xv \defineas (x_2,\dots,x_9).
\ee
Here $\Omega\in\Spin{8}$ is a rotation matrix.
We will denote this space by $\spX(R,\Omega)$.
\be\label{spXdef}
\spX(R,\Omega) = \MR{9,1}/
(x_0, x_1,\, \xv) \sim 
(x_0,x_1 + 2\pi R,\,\Omega\xv)
\ee
A field $\phi(x_0,x_1,\xv)$ in the geometry given by (\ref{sotw})
can be expanded as
$$
\phi(x_0,x_1,\xv) =
\sum_{n,\a} C_{n,\a}(x_0)
 e^{\frac{i(n-\omega_\a) x_1}{R}} Y_\a(\xv).
$$
Here $Y_\a(\xv)$ is an eigenfunction of $\Omega$ with eigenvalue $\omega_\a$,
$$
Y_\a(\Omega\xv) = e^{2\pi i\omega_\a} Y_\a(\xv).
$$
This expansion shows that states with a given
$\Omega$-charge have fractional momentum in
the $1^{st}$ direction. Similarly, in the geometry (\ref{tdtw})
a state $\ket{\psi}$ with a specific $\Omega_j$ ($j=1\dots d$) charge given
by $\omega_j$  (that is $\Omega_j\ket{\psi} = e^{2\pi i\omega_j}\ket{\psi}$)
has fractional Kaluza-Klein momentum in the $x_j$
direction and the fractional part is given by $\omega_j$.

\subsection{T-duals and U-duals of twists}\label{tudual}
We will now define a T-dual twist as follows.
By definition, type-IIB on a circle of radius $R$ with a T-dual
twist given by $\Omega$ is 
type-IIA on $\spX(\frac{\a'}{R},\Omega)$ [defined in (\ref{spXdef})].
Intuitively, we can think of the dual twist as a setting
where a state $\ket{\psi}$
with a specific $\Omega$ charge given
by $\omega$  (that is $\Omega\ket{\psi} = e^{2\pi i\omega}\ket{\psi}$)
has fractional string winding number (around the circle of radius $R$)
and the fractional part is given by $\omega$.

In a similar fashion we can define the U-duals of a twist.
We will identify the various twists according to the objects whose
charges become fractional.
Thus, we will call the T-dual twists F1-twists (F1 stands for
the fundamental string).
We can define D1-twists as the S-dual of type-IIB with an F1-twist.
We can also deform type-II string theory on $T^d$ to include
D$p$-twists. M-theory on $T^2$ can have an M2-twist parameterized
by an element of the transverse rotation group $\Omega\in\Spin{8}$.
It means that states with transverse angular momentum that transform
nontrivially under $\Omega$ have a fractional wrapped M2-brane charge.

Note that even though we have used charges to characterize
the twists we do not really need to single out the time direction.
For example, we can define a Euclidean type-IIA theory on
$\MR{10}$ with a D0-brane twist in $\Spin{10}$ as a Euclidean
version of M-theory compactified on $S^{1}$ with a geometrical twist.
This is the Melvin solution \cite{Melvin:1963qx,Costa:2000nw}.
U-duality then suggests that other D$p$-brane twists are parameterized
by elements in $\Spin{10-p}$ rather than $\Spin{9-p}$.
Similarly, in M-theory we can have M5-brane twists parameterized
by an element of $\Spin{6}$ [rather than $\Spin{5}$].

In all those cases where we have M-theory or string theory
on $T^p$ with volume $V$ and wrapped $p$-brane twists 
$\Omega\in \Spin{k}$ ($k\le 10-p$ in string theory and
 $\le 11-p$ in M-theory) we can take
the decompactification limit
$$
V\rightarrow\infty,\quad\Omega\rightarrow I,\quad
-\frac{i}{4\pi^2} V(\Omega-I)\rightarrow\hM = {\mbox{(fixed)}}\in so(k).
$$
We get a background
where an eigenstate  $\ket{\psi}$ of $\hM$ [acting as an $so(k)$
angular momentum generator] that satisfies 
$\hM\ket{\psi} = \valQ\ket{\psi}$ formally has the charge of a piece
of a $p$-brane with volume $2\pi\valQ$. (The $2\pi$ factors will simplify
upcoming formulas.)

In this way we can obtain backgrounds that are parameterized
by an element $\hM\in so(k)$ and preserve translational invariance
in $10-k$ (in string theory, or $11-k$ in M-theory) directions.

\subsection{Notation}\label{subsec:notation}
We will start from type-II toroidal compactifications where space-time is 
of the form $\R^{1,d}\times S^1\times\cdots\times S^1$.
We use the following notation for charges:
\vskip 0.5cm
\begin{tabular}{ll}
$\QD{}$ & the charge of a D$0$-brane.\\
$\QD{I_1 I_2\dots I_p}$ & the charge of a D$p$-brane that wraps directions 
                   $I_1\dots I_p$. \\
$\QKK{I}$ & the charge of a Kaluza-Klein particle in direction $I$. \\
$\QFS{I}$ & the charge of a fundamental string wrapping direction $I$. \\
$\QNS{I_1\dots I_5}$ & the charge of an NS5-brane wrapping directions
         $I_1\dots I_5$. \\
$\QM{2}{I_1 I_2}$ & the charge of an M2-brane in directions $I_1 I_2$. \\
$\QM{5}{I_1 \dots I_5}$ & the charge of an M5-brane in 
directions $I_1 \dots I_5$. \\
\end{tabular}
\vskip 0.5cm
If $i_1, i_2$ are noncompact directions,
we will denote by $\QJ{i_1 i_2}$ the angular momentum that corresponds
to rotations in the $i_1-i_2$ plane. $\QJ{i_1 i_2}$ is therefore integral
for bosons and half-integral for fermions.
A geometrical twist such as (\ref{sotw}) with $\Omega$ a rotation in the 
$2-3$ plane by an angle $2\pi\a$, for example, will be described by the
equation: 
$$
\QKK{1}-\a\QJ{2,3}\in\Z.
$$
We define
\bear
\VD{I_1 I_2\dots I_p} &\defineas& 
(2\pi)^p R_{I_1}\cdots R_{I_p}\;\QD{I_1 I_2\dots I_p},
\qquad 
\VFS{I} \defineas 2\pi R_I\;\QFS{I},
\nn\\
\VM{2}{I_1 I_2} &\defineas&
(2\pi)^2 R_{I_1} R_{I_2}\;\QM{2}{I_1 I_2},\qquad
\VM{5}{I_1\dots I_5}\defineas
(2\pi)^5 R_{I_1}\cdots R_{I_5}\;
\QM{5}{I_1 \dots I_5}
\nn\\
\VNS{I_1\dots I_5} &\defineas&
(2\pi)^5 R_{I_1}\cdots R_{I_5}\;\QNS{I_1\dots I_5},
\nn
\eear
The prefactors are the volumes of the corresponding branes.

\section{Supergravity solutions of generalized twisted backgrounds}
\label{sec:sugra}
In this section we shall study the general twisted background in 
type II string theories and M-theory. These can be obtained from
compactification of string/M-theory on a torus with nontrivial
Wilson loops for the transverse rotation group
as we demonstrated in the previous section. Probing this
background with a brane would lead to a nonlocal field theory.
One can also make a boost in  the obtained backgrounds to find backgrounds
with lightlike twist (see \cite{Simon:2002cf}).
Probing this background will also give a field theory with nonlocality
in a lightlike direction.
A special case of a D3-brane has recently been studied
in \cite{Ganor:2002ju} where a (nonlocal) dipole theory is obtained.
It is the aim of this section to generalize
this construction for other brane backgrounds in string/M-theory.

\subsection{Fundamental string twist}\label{subsec:funtw}
Concentrating on the geometry of (\ref{sotw}) we take the limit
$$
R\rightarrow 0,\qquad \Omega\rightarrow I,
\qquad
R^{-1}(\Omega-I)\rightarrow 2\pi i{\a'}^{-1}\hM = {\mbox{fixed.}}
$$
Here $\hM$ is an $so(8)$ lie-algebra valued element with dimensions
of length.
Using T-duality, the metric (in string units) is found to be
\bear
    ds^{2} &=& dt^{2} - \frac{1}{1 + {\a'}^{-2}\xvt\hM^{\top}\hM\xv}dx_1^2
        - d\xvt d\xv +\frac{(d\xvt\hM\xv)^{2}}
        {{\a'}^2+\xvt\hM^{\top}\hM\xv}
\label{wmet}
\eear
Here $\xv$ denotes the coordinate in the $8$ transverse directions
and $0\le x_1\le 2\pi$.
We also have an NSNS 2-form and a dilaton
\be\label{Bphi}
B = \frac{d\xvt\hM\xv}{{\a'}^{2}+\xvt\hM^\top\hM\xv}\wdg dx^1,
\qquad
 e^{2(\phi-\phi_0)} = \frac{1}{1 +{\a'}^{-2}\xvt\hM^\top\hM\xv}.
\ee
This is the background that we denoted by
$$
\VFS{1}-2\pi\hM_{ij}\QJ{ij}\in\Z.
$$
Note that the dilaton can be made small everywhere.
The metric, however, becomes singular as $|\xv|\rightarrow\infty$
and therefore $\a'$-corrections are important when 
$\hM |\xv|\sim \a'$.

Similarly one can find a twisted background in which $\hM$ takes its value in 
the $so(9-p)$ Lie-algebra. In this case the metric is given by
$$
ds^{2} = dt^{2} 
-\frac{1}{1 + {\a'}^{-2}\xvt\hM^{\top}\hM\xv}dx_p^2-\sum_{i=1}^{p-1}
dx_i^2 - d\xvt d\xv +\frac{(d\xvt\hM\xv)^{2}}
{{\a'}^2 + \xvt \hM^{\top}\hM\xv}\;,
$$
while the B field and the dilaton are the same as
(\ref{Bphi}).

\subsection{Generalized twists}
The S-dual configuration to (\ref{wmet})-(\ref{Bphi}) is given by
\bear
ds^{2} &=& 
e^{\phi_0}(1 + {\a'}^{-2}\xvt \hM^{\top}\hM\xv)^{\frac{1}{2}}
\left\lbrack
dt^{2} -d\xvt d\xv
\right\rbrack
\nn\\ && -e^{\phi_0}(1 + {\a'}^{-2}\xvt \hM^{\top}\hM\xv)^{-\frac{1}{2}}
\left\lbrack
dx_1^2 -{\a'}^{-2}(d\xvt\hM\xv)^{2}
\right\rbrack,
\label{Swmet}\\
C^{(RR)} &=& 
  \frac{d\xvt\hM\xv}{{\a'}^2 + \xvt\hM^\top\hM\xv}\wdg dx^1,
\quad
 e^{\phi} = e^{\phi_0}(1 +{\a'}^{-2}\xvt\hM^\top\hM\xv)^{\frac{1}{2}}
\label{SBphi}
\eear
This is the background that we denoted by
$$
\VD{1}-2\pi\hM_{ij}\QJ{ij}\in\Z.
$$
Here $\hM\in so(8)$ [or even $so(8,1)$] but if we take 
$p> 1$ and restrict
$\hM$ to an $so(9-p)$ subgroup [or perhaps $so(9-p,1)$] we
can compactify $(p-1)$ directions and apply T-duality to get
the type-II metric
\bear
ds^{2} &=& 
e^{\phi_0}(1 +{\a'}^{-(p+1)} \xvt\hM^{\top}\hM\xv)^{\frac{1}{2}}
\left\lbrack
dt^{2}
-d\xvt d\xv
\right\rbrack
\nn\\ &&
-e^{\phi_0}(1 +{\a'}^{-(p+1)} \xvt \hM^{\top}\hM\xv)^{-\frac{1}{2}}
\left\lbrack
\sum_{i=1}^p dx_i^2
-{\a'}^{-(p+1)}(d\xvt\hM\xv)^{2}
\right\rbrack,
\label{TSwmet}\\
C^{(RR)}&=& e^{\frac{p-1}{2}\phi_0}
  \frac{d\xvt\hM\xv}{{\a'}^{p+1} + \xvt\hM^\top\hM\xv}\wdg dx^1
\wdg\cdots\wdg dx^p,
\;\;\;
 e^{\phi}=\frac{e^{\phi_0}}{
   (1 +{\a'}^{-(p+1)}\xvt\hM^\top\hM\xv)^{\frac{p-3}{4}}}.
\nn\\ &&
\label{TSBphi}
\eear
In the above formula we have absorbed a factor of ${\a'}^{\frac{p-1}{2}}$
in $\hM$ so as to make it of dimensions $[{\mbox{length}}]^p$.
In the notation of \ref{subsec:notation} this background corresponds to
$$
\VD{1\dots p}-2\pi\hM_{ij}\QJ{ij}\in\Z.
$$
We can lift the type-IIA metric with a D2-brane twist to obtain
M-theory with an M2-twist.
\bear
ds^{2} &=& 
(1 + l_p^{-6}\xvt\hM^{\top}\hM\xv)^{\frac{1}{3}}
\left\lbrack
dt^{2} -d\xvt d\xv
-dx_{10}^2
\right\rbrack
\nn\\ &&
-(1 + l_p^{-6}\xvt\hM^{\top}\hM\xv)^{-\frac{2}{3}}
\left\lbrack
\sum_{i=1}^2 dx_i^2 -l_p^{-6}(d\xvt\hM\xv)^{2}
\right\rbrack,
\label{Mwmet}\\
C &=& \frac{d\xvt\hM\xv}{l_p^6 + \xvt\hM^\top\hM\xv}\wdg
  dx^1\wdg dx^2,
\label{MCfield}
\eear
In the notation of \ref{subsec:notation} this background corresponds to
$$
\VM{2}{12}-2\pi\hM_{ij}\QJ{ij}\in\Z.
$$
We can also lift the type-IIA metric with a D4-brane twist to
obtain M-theory with an M5-twist.
\bear
ds^{2} &=& 
(1 + l_p^{-12}\xvt\hM^{\top}\hM\xv)^{\frac{2}{3}}
\left\lbrack
dt^{2} -d\xvt d\xv
\right\rbrack 
\nn\\ &&
-(1 + l_p^{-12}\xvt\hM^{\top}\hM\xv)^{-\frac{1}{3}}
\left\lbrack
\sum_{i=1}^4 dx_i^2 +dx_{10}^2
-l_p^{-12}(d\xvt\hM\xv)^{2}
\right\rbrack,
\label{MwmetF}\\
{}^{*}dC &=& 
\left\lbrack
\frac{d\xvt\wdg\hM d\xv}{l_p^{12} + \xvt\hM^\top\hM\xv}
-2\frac{d\xvt\hM\xv\wdg d\xvt\hM^\top\hM\xv}{(l_p^{12} + \xvt\hM^\top\hM\xv)^2}
\right\rbrack
    \wdg dx^1\wdg\cdots\wdg dx^4\wdg dx^{10},\nn\\ &&
\label{MCfieldF}
\eear
In the notation of \ref{subsec:notation} this background corresponds to
$$
\VM{5}{1,\dots, 4,10}-\hM_{ij}\QJ{ij}\in\Z.
$$
Finally we can lift the type-IIA metric with a D6-brane twist.
In this case $\xv$ is 3-dimensional and we can take $\hM$ to be
proportional to the generator of rotations in directions $8,9$.
Let ${\a'}^{\frac{7}{2}}\mQ$ denote its magnitude.
We also set $z\defineas x_8 + i x_9$.
Thus,
$$
1 + {\a'}^{-(p+1)}\xvt \hM^{\top}\hM\xv= 1 + \mQ^2|z|^2,\qquad
d\xvt\hM\xv = \frac{i\mQ}{2}{\a'}^{\frac{7}{2}}(z d\bz -\bz dz).
$$
\bear
    ds^{2} &=& 
e^{\phi_0}(1 + \mQ^2|z|^2)^{\frac{1}{2}}
\left\lbrack
dt^{2} -|dz|^2 -dx_7^2
\right\rbrack
\nn\\ &&
-e^{\phi_0}(1 + \mQ^2|z|^2)^{-\frac{1}{2}}
\left\lbrack
\sum_{i=1}^6 dx_i^2
-\frac{\mQ^2}{4}|z d\bz -\bz dz|^2
\right\rbrack,
\label{TSwmetII}\\
dC^{(RR)} &=& 
\frac{i {\a'}^{-\frac{7}{2}}e^{\frac{5}{2}\phi_0}\mQ}{
   (1 + \mQ^2|z|^2)^2}dz\wdg d\bz\wdg 
   dx^1\wdg\cdots\wdg dx^6,
\quad
 e^{\phi} = e^{\phi_0}(1 + \mQ^2|z|^2)^{-\frac{3}{4}}.
\nn\\ &&
\label{TSBphiII}
\eear
Note that
$$
\sqrt{-g} = (1 + \mQ^2|z|^2)^{-1}.
$$
The dual RR field is
$$
{}^{*}dC^{(RR)} = l_p^{-1}\mQ dt\wdg dx^7.
$$
Thus the 1-form RR-field can be taken to be
$$
A^{(RR)} = -l_p^{-1}\mQ x^7 dt.
$$
Lifting to M-theory and setting $z=r e^{i\theta}$ we obtain the metric
\bear
ds^{2} &=& 
(1 + \mQ^2 r^2)\left\lbrack dt^{2} -dr^2 -dx_7^2 \right\rbrack
-r^2 d\theta^2 -\frac{1}{1 + \mQ^2 r^2}(dx_{10}+\mQ x_7 dt)^2
 -\sum_{i=1}^6 dx_i^2
\nn\\ &&
\label{MKKMwmet}
\eear
and no fluxes.
But note that it is not supersymmetric and the issue of stability is
not clear.

\subsection{Lightlike twists}
We can also find lightlike twisted backgrounds of string theory or M-theory 
by making use of a boost from the dipole twisted backgrounds which have
been described so far (see also \cite{Figueroa-O'Farrill:2001nx}). 
This can also be thought of as taking the Penrose 
limit of the corresponding dipole twisted backgrounds. 
To begin with we consider the following  boost
in the $x_p$ direction  
\be
{\hat t}=\cosh\gamma\;t-\sinh\gamma\;x_p,\;\;\;\;\;
{\hat x}_p=-\sinh\gamma\;t-\cosh\gamma\;x_p\;,
\label{BBOOST1}
\ee
or 
\be
x^+=e^{-\gamma}y^+,\;\;\;\;\;x^{-}=e^{\gamma}y^-\;,
\label{BBOOST2}
\ee
with $y^{\pm}=x_p\pm t$ and $x^{\pm}={\hat x}_p\pm {\hat t}$.

To have a lightlike dipole we now take the infinite boost limit,
$\gamma \rightarrow \infty$.
In order to end up with a lightlike dipole vector with finite component
we must simultaneously scale $\hM\rightarrow 0$ while  keeping the 
following quantity fixed
\be
\tM\defineas \hM e^{\gamma}={\rm finite}
\label{BLLDI}
\ee
In this limit the background (\ref{TSBphi}) reads
\bear
ds^{2} &=& dx^{+} dx^{-} 
+{1\over 4}{\a'}^{-(p+1)}(\xvt\tM^{\top}\tM\xv)(dx^{+})^2 
 -d\xvt d\xv -\sum_{i=1}^{p-1}(dx^i)^2,
\nn\\
C^{RR} &=& {1\over 2}{\a'}^{-(p+1)} 
(d\xvt\tM\xv) \wdg dx^1\wdg\cdots\wdg dx^{p-1}\wdg dx^{+},
\qquad
 e^{2(\phi-\phi_{0})} = 1,
\eear
which is the RR pp-wave specified by
$$
\VD{1,\cdots, (p-1), +}-2\pi\tM_{ij}\QJ{ij}\in\Z\, .
$$
Similarly we can also find the NSNS pp-wave background by making use of 
a boost from the twisted backgrounds studied
in subsection \ref{subsec:funtw}.  
For example the lightlike background specified by
$$
\VFS{+}-2\pi\tM_{ij}\QJ{ij}\in\Z,
$$
is the NSNS pp-wave given by
\bear
ds^{2} &=& dx^{+} dx^{-} 
+{1\over 4}\;{\a'}^{-2}(\xvt\tM^{\top}\tM\xv)(dx^{+})^2 - d\xvt d\xv,
\nn\\
B &=& {1\over 2}\;{\a'}^{-2}(d\xvt\tM\xv) \wdg dx^{+},
\qquad
 e^{2(\phi-\phi_{0})} = 1.
\eear
The same procedure can be applied to M-theory twisted backgrounds. 
In this way we will be able to find the lightlike twisted background in
M-theory.
For example from the M2-twist background (\ref{Mwmet}) of M-theory using 
a boost similar to (\ref{BBOOST1}) and (\ref{BBOOST2}) one finds
\bear
ds^2 &=& dx^{+}dx^{-} 
+{1\over 4}\; l_p^{-6}(\xvt\tM^\top\tM\xv)(dx^{+})^2 - 
d\xvt d\xv -dx_1^2-dx_{10}^2\, ,\cr
C&=&{1\over 2}\; l_p^{-6}(d\xvt\tM\xv)\wdg dx^1\wdg dx^{+}\;,
\eear
which corresponds to
$$
\VM{2}{1+} -2\pi\tM_{ij}\QJ{ij}\in\Z.
$$
The lightlike twist background of M-theory corresponding to
$\VM{5}{1,\cdots,4,+}-2\pi\tM_{ij}\QJ{ij}\in\Z$ can also be obtained from
the Penrose limit of the M5-twist background. 

We note that the pp-wave backgrounds studied in this section provide
string theory backgrounds in which the string theory can be exactly
solved. The lightlike twist will also provide pp-wave backgrounds of 
M-theory and it would be interesting to study the Matrix model of these
pp-wave backgrounds. In particular one can find a pp-wave-like background in
M-theory without any fluxes. This background can be obtained by taking
the Penrose limit from (\ref{MKKMwmet})
\bear
ds^2&=&dx^{+}dx^{-} +\frac{1}{4}\tM^2|z|^2 (dx^{+})^2-|dz|^2-dx_7^2
-\sum_{i=1}^{5}dx_i^2
-\left(dx_{10}+\frac{1}{2}\tM x^7 dx^{+}\right)^2\;.
\nn\\ &&
\eear
We note, however, that it is not obvious whether this is a stable background
of M-theory.
Backgrounds somewhat reminiscent of this have recently been discussed
in \cite{Grandi:2002gt}-\cite{Herdeiro:2002ft}.
 
\section{Probing with branes and new nonlocal theories}
\label{sec:probes}
We will now examine brane probes in the various twisted geometries.
In many cases we will discover new types of nonlocal field theories.
The configurations that we will discuss are related to configurations
of D-branes in pp-wave backgrounds.
Such configurations have been discussed extensively in
\cite{Michishita:2002jp}\cite{Balasubramanian:2002sa}-\cite{Metsaev:2002sg}
but mostly in the context of the pp-wave limit
relevant for $AdS_5\times S^5$ \cite{Berenstein:2002jq}
and not so much for the twisted backgrounds.

\subsection{Fundamental string twists}\label{FunTw}
In this case $p=1$ and $\hM$ has dimensions of length.
We take the fundamental string twist to be in the direction of $x_1$
and take $N$ D$d$-branes that extend in directions $0,1\dots d$.
The twist $\hM$ is then an element of the Lie algebra $so(9-d)$
corresponding to rotations in the remaining directions.
This case of a fundamental string twist has been studied
\cite{Bergman:2001rw,Alishahiha:2002ex}
and in the case that
$\hM \gg {\a'}^{1/2}$ the resulting low-energy 
description of the dynamics of the brane is a theory of fundamental
dipoles with lengths proportional to the eigenvalues of $2\pi\hM$.
In order to be self-contained we have included a review of dipole theories
in appendix \ref{app:review}.
The proof that the dipole-theory is indeed the low energy description
on the brane probe will not be repeated here,
but intuitively we can argue as follows.
On a D-brane the fundamental string is identified with electric flux.
So dipoles naturally behave like fixed size open strings. The statement
that states with $\Spin{9-d}$ charge also have a finite string length
translates on the D-branes to the statement that states with
R-symmetry charge are dipoles.

If, on the other hand,
the D-brane probes are transverse to the twist we get a massive
deformation of Super Yang-Mills theory.
For example, let us take $N$ D$d$-probes in directions $0,2,\dots,d$.
It is convenient to compactify the direction of $x_1$ on a circle
of radius $R\gg |\hM|$.
T-duality on that circle will give us D$d$-branes wrapped
on a circle of radius $\a'/R$ with a geometrical twist of 
magnitude $\frac{\hM}{R}$.
It is not hard to see that the effect of the twist on the 
low energy description in $d$ dimensions is to add 
a mass term to the scalars and fermions of $d$ dimensional
$U(N)$ Super Yang-Mills theory. The mass term is given by
\be\label{MassFtw}
4\pi^2\a'^{-2}\phi^\top\hM^\top\hM\phi +2\pi\a'^{-1}\psi^\top\hM\psi.
\ee
Here $\phi$ represents the scalars, written as a vector in the 
fundamental representation of $so(9-d)$ and 
$\psi$ represents the fermions, written as a vector in the spinor 
representation of $so(9-d)$. $\hM$ that appears in the formula above
should be interpreted as  a matrix in the appropriate representation.

Again, we can heuristically understand the mass term as follows.
In the presence of the twist, particles with R-symmetry
charge behave like finite fundamental strings that
are perpendicular to the D-branes and have length
proportional to the corresponding eigenvalue of $2\pi\hM$.
Their mass is therefore proportional to $2\pi\a'^{-1}\hM$.
Of course, in string theory there are no finite fundamental strings
perpendicular to the D-brane but the intuitive picture gives the correct
mass for the R-symmetry charged particles.
Using the supergravity solution
(\ref{wmet})-(\ref{Bphi}) the mass term can be interpreted as 
a gravitational potential that attracts the D-brane probe to the
origin of the transverse space $\MR{9-d}$ \cite{Chakravarty:2000qd}.

\subsection{D-brane twists}\label{subsec:Dbrtwists}
We get new nonlocal field theories
when we probe the {\it generalized} twisted backgrounds with D-branes.
For most of these cases, we do not have a simple field theory 
description.
Let us explore various new possibilities that arise.


\subsubsection*{D3-probes with a longitudinal D1-twist}
This is the S-dual theory to the D3-probes with a 
longitudinal fundamental string twist.
Since the latter is described by the dipole-theory
the S-dual should be a field theory with fundamental magnetic dipoles.
For a small dipole vector matrix $\hM$ the electric dipole theories
can be described \cite{Bergman:2000cw}
as a deformation of $\SUSY{4}$ $U(N)$ SYM by the operator
$$
2\pi\hM_{IJ}^{\u} \cO^{IJ}_\u,\qquad (I,J=1\dots 6),\qquad
(\u=0\dots 3)
$$
where $\cO^{IJ}_\u$ is the dimension-$5$ operator
\bear
\cO_\u^{IJ} &=&
       {i\over {\gYM^2}}
       \tr{{F_{\u}}^\v\Phi^{[I} D_\v\Phi^{J]}
   +\sum_{K}(D_\u\Phi^K)\Phi^{[K}\Phi^I\Phi^{J]}}
        + {\mbox{fermions}} \nn
\eear
Here $I,J=1\dots 6$ are R-symmetry indices,
$\Phi^I$ ($I=1\dots 6$) are the scalars,
$D_\u = \px{\u} -i [A_\u,\cdot]$ is the
covariant derivative, $F_{\u\v}$ is the field strength
and $[\cdots ]$ means complete anti-symmetrization.
${\cal O}_\u^{IJ}$
is a vector operator that transforms in the $\rep{15}$ of the
R-symmetry group $SU(4)$.

The magnetic dipole theory should be described, to linear order
in $\hM$, by the deformation $2\pi\hM_{IJ}^{\u} \cwO^{IJ}_\u$
where $\cwO^{IJ}_\u$ is the dual dimension-$5$ operator
\bear
\cwO_\u^{IJ} &=&
       {i\over {\gYM^2}}
       \tr{{\wF_{\u}}^\v\Phi^{[I} D_\v\Phi^{J]}
   +\sum_{K}(D_\u\Phi^K)\Phi^{[K}\Phi^I\Phi^{J]}}
        + {\mbox{fermions}} \nn
\eear
where $\wF_{\u\v} = \frac{1}{2}\epsilon_{\u\v\sigma\tau} F^{\sigma\tau}$
is the dual (nonabelian) field strength.

\section{Supergravity solutions of D-branes and NS5-branes and
the large $N$ limit}
\label{sec:LargeN}
In this section we shall first review Dp-brane probes of the F1 twist 
geometry. The obtained supergravity solution is a Dp-brane solution 
in  the presence of B field with one leg along the worldvolume and the 
other along the directions transverse to the brane. 
This background would provide the supergravity description of noncommutative
dipole field theory\footnote{Special cases of dipole theories have been discussed
in \cite{{Cheung:1998te}, {Banks:1999tr},{Dasgupta:2000ry}} and various aspects 
of the theories have been explored in \cite{Motl:2001dj}-\cite{Huang:2002yg},
see also \cite{Alishahiha:2002ex}.}. 
We then consider a 
new set of nonlocal theories by performing a boost along the worldvolume 
direction of the brane in which the $B$ field is nonzero.
Generalizations to
NS5-branes with RR field and M-theory with $C$ form
field will also be studied. 

We note also that making a boost (the Penrose limit) in a nonlocal theory has 
recently been considered in \cite{Hubeny:2002vf}-\cite{Bhattacharya:2002qx}.

\subsection{Review of the supergravity dual of dipole theory} 
By probing the F1 twist geometry with Dp-branes we find a 
supergravity solution of Dp-branes in the presence of a B-field 
with one leg along the worldvolume and the other along the transverse 
directions to the brane \cite{Alishahiha:2002ex}
\bear
ds^2 &=& f^{-\frac{1}{2}} \left( dt^2 - dx^2_1 -\cdots-dx^2_{p-1} 
- \frac{ \alpha'^2 d x^2_p}{ \alpha'^2 + r^2 \nvt \hMt \hM \nv} \right) 
\cr && \cr
&-& f^{\frac{1}{2}} \left( dr^2 + r^2 d\nvt d\nv - 
\frac{r^4 (\nvt \hMt d\nv)^2}{\alpha'^2 + r^2 \nvt \hMt \hM \nv} 
 \right)\;, \cr &&\cr 
 e^{2\phi} &=&  \frac{\alpha'^2 g^2_s f^{\frac{3-p}{2}}}{{\a'}^2 +
 r^2 \nvt \hMt \hM \nv }\; ,\cr &&\cr 
\sum_{a=p+1}^9 B_{pa} dx_a &=&  \frac{r^2 d\nvt \hM \nv}{ \alpha'^2  + 
r^2 \nvt \hMt \hM \nv}\;,
\label{GenSol}
\eear
where $\nv$ is an $(8-p)$-dimensional unit vector, $|\nv|^2=1$, and
$$
f =1+ \frac{(4\pi)^{\frac{5-p}{2}}\Gamma\left(\frac{7-p}{2}\right) 
N g_s {\a'}^{\frac{7-p}{2}}}{r^{7-p}}.
$$
Also, $\hM$ is the same $so(8-p)$ element from
subsection \ref{subsec:funtw}.
In general this background breaks the supersymmetry completely, 
but for special cases of the matrix $\hM$
some supersymmetries can be left intact.
Another problem that has to be considered is the stability 
of the solution. It is not obvious that the solutions we found are 
stable. Nevertheless, taking  the matrix $\hM$ such that the solutions 
preserve some  amount of supersymmetries, as we will consider, 
would hopefully lead to stable solutions. In fact, in this paper we 
shall mostly 
consider matrices $\hM$ such that 8 supersymmetries are preserved.

It has also been argued \cite{Alishahiha:2002ex}
that the worldvolume theory of the 
supergravity solution (\ref{GenSol}) decouples from bulk gravity
leading to a nonlocal theory, i.e. noncommutative dipole 
theory (reviewed in appendix \ref{app:review}).
The decoupling limit, in which the worldvolume theory
decouples from the bulk, is defined by $\alpha'\rightarrow 0$
keeping the following quantities fixed
\be
u\defineas {r\over \a'}\;,\;\;\;\;\;\;
\bgs\defineas {\a'}^{\frac{p-3}{2}} g_s\; .
\ee

In this limit the supergravity solution (\ref{GenSol}) reads
\bear
{\a'}^{-1} ds^2 &=& \left(\frac{u}{R}\right)^{\frac{7-p}{2}}
\left( dt^2 - dx_1^2 -\cdots- \frac{dx_p^2}{1+u^2 \nvt \hMt \hM \nv} \right)
 \cr &&\cr &-& \left(\frac{R}{u}\right)^{\frac{7-p}{2}} 
 \left(du^2 + u^2 d\nvt d\nv 
- \frac{u^4(\nvt \hMt d\nv)^2}{1+u^2 \nvt \hMt \hM \nv} \right)\; , \cr &&\cr
e^{2 \phi} &=& \bgs^2 
\frac{(\frac{R}{u})^{(7-p)(3-p)/2}}{1+u^2 \nvt \hMt \hM \nv}
\;,\cr &&\cr
\sum_{a=p+1}^9B_{pa}d\nv_a &=&  
\frac{u^2 d\nvt\hM \nv}{1 + u^2 \nvt \hMt \hM \nv}\; ,
\label{NEAR}
\eear
with
\be
R^{7-p} = 2^{7-2p} \pi^{\frac{9-3p}{2}} \Gamma
\left(\frac{7-p}{2}\right)\gYM^2 N\;,
 \;\;\;\;\;\;
\gYM^2 = (2\pi)^{p-2} \bgs \; .
\ee
This supergravity solution provides the gravity dual of the noncommutative
dipole gauge theory.

Starting from the case of $p=5$ and applying S-duality one can find the
supergravity solution of type IIB NS5-branes in the 
presence of an RR 2-form potential with one leg along the brane and the
other along the transverse directions. This could provide the gravity
dual of the dipole deformation of little string theory. We can also make
a series of T-duality transformations
to produce a new supergravity solution. This supergravity
solution describes type II NS5-branes 
in the presence of RR $(6-p)$-form, for $p=0\dots 4$, with
one leg along the transverse directions and $(5-p)$ legs along
the NS5-branes worldvolume. The corresponding supergravity solution
in the decoupling limit is given by
\bear
ds^2&=&(1+u^2L^2)^{1/2}\bigg{[}dt^2-\sum_{i=1}^{p}dx_i^2-
\frac{\sum_{j=p+1}^5dx_j^2}
{1+u^2L^2} \cr &&\cr 
&-&{N{\a'}\over u^2}\bigg{(}du^2+u^2d\Omega_3-\frac{u^4L^2}
{1+u^2L^2}(a_1d\theta_1+a_2d\theta_2  
+  a_3d\theta_3)^2\bigg{)}\bigg{]}\;,
\cr &&\cr
e^{2\phi}&=&\frac{N}{l_s^2 u^2}(1+u^2L^2)^{(p-2)/2}\;,\cr &&\cr
\sum_{a=6}^{9}C_{(p+1)\cdots 5\theta_a}d\theta_a&=&\frac{u^2L}
{1+u^2L^2}(a_1d\theta_1+a_2d\theta_2+a_3d\theta_3)\;,
\label{SUGNS5}
\eear
where $\theta_i$'s are angular coordinates parameterizing the sphere 
$S^{3}$ transverse to the NS5-branes, and
\be
a_1\defineas \cos\theta_2,\;\;\;\;\;
a_2\defineas -\sin \theta_1 \cos \theta_1 \sin \theta_2,\;\;\;\;\;
a_3\defineas\sin^2 \theta_1 \sin^2 \theta_2\;.
\label{ASM5}
\ee
Note that in the above solution we have chosen the matrix $\hM$ in such a 
way that the system is maximally supersymmetric which means that 
the solution preserves 8 supercharges. For this case the matrix $\hM$ has the
following form
\be
\hM=\pmatrix{0&L&0&0\cr -L&0&0&0\cr 0&0&0&L\cr 0&0&-L&0}.
\ee

\subsection{Lightlike dipole theory}
In this section we shall study the lightlike dipole theory using its
gravity description. To find the corresponding supergravity 
solution we start with a background in which the probe Dp-branes have a
small spacelike dipole vector and then we perform a large boost.
In fact we will consider the following boost
in the $x_p$ direction  
\be
{\hat t}=\cosh\gamma\;t-\sinh\gamma\;x_p,\;\;\;\;\;
{\hat x}_p=-\sinh\gamma\;t-\cosh\gamma\;x_p\;,
\label{BOOST1}
\ee
or 
\be
x^+=e^{-\gamma}y^+,\;\;\;\;\;x^{-}=e^{\gamma}y^-\;,
\label{BOOST2}
\ee
with $y^{\pm}=x_p\pm t$ and $x^{\pm}={\hat x}_p\pm {\hat t}$.

To have a lightlike dipole we now take the infinite boost limit,
$\gamma \rightarrow \infty$.
In order to end up with a lightlike dipole vector with finite component
we must, at the same time, also take the limit 
$\hM\rightarrow 0$ while keeping the following quantity fixed
\be
\cM\defineas\hM e^{\gamma}={\rm finite}
\label{LLDI}
\ee
In this limit the background (\ref{NEAR}) reads
\bear
{ds^2\over l_s^2}&=&\left(\frac{u}{R}\right)^{\frac{(7-p)}{2}}\left[
-dx^+dx^-+{\nvt\cMt\cM \nv\over 4}\;u^2(dx^+)^2-
dx_1^2-\cdots-dx_{p-1}^2\right]\cr &&\cr
&-& \left(\frac{R}{u}\right)^{\frac{7-p}{2}}\left[du^2+u^2d\Omega_{8-p}^2
\right]\;,\cr &&\cr
e^{2\phi}&=&\bgs^2\left({R\over u}\right)^{{
(7-p)(3-p)\over 2}}, 
\label{lltwmet}
\eear
and
\be
\sum_{a=p+1}^{9}B_{+a}d\nv_a={1\over 2}u^2 d\nvt \cM \nv\;.
\label{llBtw}
\ee
One can now choose the matrix $\hM$ such that the solution preserves 
8 supercharges. For this Dp-brane case such a matrix $\hM$ is given by
\be
\cM=L(e_{6-p,7-p} -e_{7-p,6-p} +e_{8-p,9-p}-e_{9-p,8-p})\;,
\ee
where $e_{ij}$ are a set of $(9-p)^2$ matrices
of dimensions $(9-p)\times (9-p)$
and are defined by $(e_{ij})_{kl}=\delta_{ik} \delta_{jl}$.
The effective dipole moment of the noncommutative dipole 
theory described by (\ref{NEAR}) is defined to be
\be
2\pi\Leff\defineas 2\pi L\prod_{i=1}^{5-p}\sin\theta_i,\;\;\;\;\;\;
{\rm for}\;p=1,\cdots,5\;,
\label{effdipole}
\ee
where $\theta_j$ [$j=1\dots (8-p)$] are angular coordinates
parameterizing the sphere $S^{(8-p)}$ transverse to the Dp-brane such
that
$$
\nv = (
\cos\theta_1,\,
\sin\theta_1\cos\theta_2,\,
\sin\theta_1\sin\theta_2\cos\theta_3,\,
\cdots,\,
\prod_{j=1}^{8-p}\sin\theta_j).
$$
In the large boost limit we have $L\rightarrow 0$ while
$e^{\gamma}L$ is kept fixed.
We define
\be\label{mueffdef}
2\pi\mueff(\theta_1,\dots,\theta_{8-p})\defineas 2\pi e^{\gamma}\Leff
\ee
to be the finite effective magnitude of the lightlike dipole vector.

For this case with 8 preserved supercharges
the metric (\ref{lltwmet}) reads
\bear
{\a'}^{-1}ds^2&=&\left(\frac{u}{R}\right)^{\frac{(7-p)}{2}}\left[
-dx^+dx^-+{\mueff^2 \over 4}\;u^2(dx^+)^2-
dx_1^2-\cdots-dx_{p-1}^2\right]\cr &&\cr
&-& \left(\frac{R}{u}\right)^{\frac{7-p}{2}}\left[du^2+u^2d\Omega_{8-p}^2
\right]\;, 
\eear

\subsection{Deformations of little string theory}
One can also proceed to study the lightlike deformation of 
little string theory. To do this, we start from the supergravity
solution (\ref{SUGNS5}) and perform a boost in the $x_5$ direction.
Using a boost similar to (\ref{BOOST1})-(\ref{BOOST2}) 
in the limit $\gamma\rightarrow \infty$ and taking into account
the finiteness condition (\ref{LLDI}) for the lightlike dipole
the supergravity solution (\ref{SUGNS5}) reads
\bear
ds^2&=&-dx^+dx^-+{\mu^2\over 4}u^2(dx^+)^2
-N{\a'}{du^2\over u^2}-N{\a'}d\Omega_3^2-\sum_{i=1}^4dx_i^2\;,\cr &&\cr
e^{2\phi}&=&{N\over {\a'}u^2},\;\;\;\;\;
\sum_aC_{(p+1)\cdots 4 +\theta_a}d\theta_a
={\mu u^2\over 2}(a_1d\theta_1+a_2d\theta_2+a_3d\theta_3),
\label{LIO}
\eear
where $\mu=e^{\gamma}L$ is again the magnitude of the lightlike dipole vector. 
Note that there is also a two form $B$ field representing the 
charge of the NS5-branes which is given by
\be
dB={\a'}N\epsilon_3\;,
\ee
where $\epsilon_3$ is the volume form of the $S^3$ sphere transverse to the 
NS5-branes.

Interestingly enough, the string theory in this obtained background 
can be exactly solved. Actually, the string theory in this  
Liouville PP-wave background in light-cone gauge can be described by 
the level $N\;$ $SU(2)$ WZW model plus a Liouville field
and four free scalars. The Liouville PP-wave background in
string theory has been considered in
\cite{{Maldacena:2002fy},{Russo:2002qj}} as a background in which the
string theory can be exactly solved. It would be interesting to analyze
the string theory in the background (\ref{LIO}).

\subsection{Supergravity duals of disc theories}
In this section we will consider a nonlocal theory where the 
parameter of nonlocality is a tensor.
This theory can be naturally defined as the worldvolume
theory of M5-branes in the presence of a 3-form $C$-field
with two legs along the worldvolume and one leg transverse
to it.\footnote{
We note also that the theory on the NS5-branes studied in the
previous section could have tensor nonlocality parameters
given by RR $n$-form field strengths for $n>3$.} To find the
corresponding supergravity solution we can start from
a $D4$ brane solution and then lift it to 11-dimensional 
supergravity and send the radius of the $11^{th}$ direction to infinity,
$R_{11}\rightarrow \infty$. In this limit, setting $R_{11}\hM=\bM$,
one finds \cite{Alishahiha:2002ex} :
\bear
ds_{11}^2&=& h^{-1/3}\bigg{[} 
f^{-1/3}\bigg{(} dt^2-\cdots-dx_3^2- h\;(dx_4^2+dx_5^2)\bigg{)}\cr&&\cr 
&+& f^{2/3}
\bigg{(}dr^2+r^2d\nvt d\nv -h\;{r^4\over l_p^6}\;(\nvt\bM d\nv)^2 
 \bigg{)}\bigg{]}\;,\cr &&\cr
\sum_{a=2}^4C_{45a}dx^a&\sim& h\;{r^2\over l_p^6}\;d\nvt\bM \nv\;,
\eear
where
\be
f\defineas 1+\frac{\pi N l_p^3}{r^3},\;\;\;\;\; h^{-1}\defineas
1+{r^2\over l_p^6}\nvt\bMt\bM\nv\;.
\ee
The decoupling limit of the theory is defined as a limit where 
$l_p\rightarrow 0$ keeping $u={r\over l_p^3}$ fixed. In this limit,
setting $\bM$ as
\be
\bM=\pmatrix{0&0&0&0&0\cr 0&0&{\bar L}&0&0\cr 0&-{\bar L}&0&0&0\cr
0&0&0&0&{\bar L}\cr 0&0&0&-{\bar L}&0}\;,
\ee 
to preserve 8 supercharges, the above supergravity solution reads
\bear
l_s^2ds^2&=& h^{-{1\over 3}}\bigg{[}\frac{u}{(\pi N)^{1/3}}
\left( dt^2 - dx_1^2 - \cdots+dx_3^2- 
h\;(dx_4^2+dx_5^2) \right) \cr &&\cr 
 &-& \frac{(\pi N)^{{2\over 3}}}{u^2} \left( du^2 + u^2 d \Omega_3^2 - 
 h u^4 \Leff^2 
(a_2d\theta_2 +a_3 d\theta_3 + a_4d \theta_4)^2 \right)\bigg{]},\cr &&\cr
\sum_{a=2}^4C_{45\theta_a}d\theta_a&=& h u^2\Leff(a_2d \theta_2 +
 a_3 d\theta_3+a_4 d\theta_4)\; ,
 \label{M5D}
\eear 
where $h^{-1}=1+u^2\Leff^2$ and
$\theta_1,\cdots,\theta_4$ are angular coordinates 
parameterizing the
sphere $S^4$ transverse to the brane and $a_i$'s are given
by 
\be
a_2\defineas \cos\theta_3,\hspace{1cm}
a_3\defineas -\sin \theta_2 \cos \theta_2\sin \theta_3,\hspace{1cm}
a_4\defineas\sin^2 \theta_2 \sin^2 \theta_3\; .
\label{AD4}
\ee
The effective ``{\it discpole}'' is also defined by 
\be\label{LeffDP}
2\pi\Leff = 2\pi {\bar L} \sin\theta_1
\ee
where ${\bar L}$ has dimension of $({\rm length})^2$.

It is also possible to find a {\it lightlike discpole} theory by making
use of a boost in the $x_5$ direction. Consider a boost in the $x_5$
direction given by (\ref{BOOST1}) and (\ref{BOOST2}). In the limit
$\gamma\rightarrow \infty$ and taking into account
the finiteness of the lightlike dipole [requirement (\ref{LLDI})], one finds
\bear
ds^2&=&{u\over (\pi N)^{1/3}}\left(-dx^+dx^-+{\mueff^2\over 4}
u^2(dx^+)^2-dx_1^2-\cdots-dx_4^2\right)\cr && \cr 
&-&{(\pi N)^{2/3}\over u^2}\;(du^2+u^2d\Omega_4^2)\;,\cr &&\cr
\sum_{a=2}^4C_{+4\theta_a}d\theta_a&=-& {\mueff \over 2}\;u^2(a_2d \theta_2 +
 a_3 d\theta_3+a_4 d\theta_4)\; ,
\label{LLDISC}
\eear
where $2\pi\mueff=2\pi e^{\gamma}\Leff$ is the finite $\theta$-dependent
magnitude of the lightlike dipole.
Note that in the large boost limit we have $L\rightarrow 0$ while
$e^{\gamma}L$ is kept fixed.

To study the nonlocal structure of the theory it would be useful to
study the expectation values of Wilson loops/surfaces in the theory. 
In the next section we will study the Wilson loops/surfaces in the
theories we have studied so far using the corresponding supergravity
solutions.

\subsection{The nondecoupling of the center $U(1)\subset U(N)$}
But before we proceed let us discuss the $U(1)$ center of the gauge group.
In the local $\SUSY{4}$ Super-Yang-Mills theory the traces
of the gauge field, scalars and fermions (viewed as $N\times N$ matrices)
decouple from the rest of the Lagrangian which describes an $SU(N)$
gauge theory. This decoupling has also been demonstrated in the 
supergravity dual \cite{Witten:1998qj,Aharony:1998qu}.
On the other hand, it is well known that in noncommutative $U(N)$ 
gauge theories the traces of the fields do not decouple. 
On the supergravity dual this has been explained in \cite{Gross:2000ba}
as a consequence of a nonzero 3-form NSNS flux and a term of
the form $\int B_2^{RR}\wedge F_3^{NS}\wedge F_5^{RR}$ in the bulk 
type-IIB supergravity.
Here $F_3$ and $F_5$ are the 3-form NSNS and 5-form RR field strengths.
In the supergravity dual of noncommutative
Yang-Mills theory \cite{Hashimoto:1999ut,Maldacena:1999mh}
the field strengths are oriented in such a way that turning on
a nonzero $B_2^{RR}$ in the direction perpendicular to the noncommutativity
costs energy and this is related to the energy of electric fluxes.

What happens in dipole theory?
The $U(1)$ gauge field does not decouple but, unlike noncommutative
Young-Mills theory, the dipole theory is well defined for an $SU(N)$
gauge group. Therefore, the question arises which gauge group does
the supergravity dual describe?
Here again there is a nonzero $F_3^{NS}$ but it has a component in the
direction of the $S^5$ and therefore if $B_2^{RR}$ has no legs in
the direction of the $S^5$ then
$B_2^{RR}\wedge F_3^{NS}\wedge F_5^{RR} = 0$.
This suggests that the supergravity dual describes the dipole theory
with $SU(N)$ rather than $U(N)$ gauge group.
This is just as well because the $U(N)$ dipole theory is probably 
ill-defined in the UV (see the $\beta$-function calculations
in \cite{Dasgupta:2001zu,Sadooghi:2002ph}).

What about the trace of the scalars and fermions?
{}In the local $\SUSY{4}$ Super-Yang-Mills it was
argued in \cite{Witten:1998qj} that operators such as
$\tr{\Phi^I}$ (using the notation of subsection \ref{subsec:Dbrtwists}),
if they existed, would correspond to supergravity fields in $AdS_5\times
S^5$ with
conformal dimension $\Delta=1$ and they would violate the
Breitenlohner-Freedman bound $\Delta \ge 2$.
As explained there, this bound is related to the fact that
if the conformal dimension is too low
the corresponding supergravity mode converges too fast near
the boundary. The metric (\ref{NEAR}) behaves very differently from
$AdS_5\times S^5$ near the boundary $u\rightarrow\infty$ and therefore
the results about $AdS_5\times S^5$ do not apply.
In fact (\ref{NEAR}) has strong curvature for large $u$ and the
supergravity approximation is not applicable near the boundary.

In the dipole theories the expression $\tr{\Phi^I}$ is not gauge invariant
because $\Phi^I(x)$ transforms as a product of a quark and anti-quark fields
at two different points. To make $\tr{\Phi^I}$ gauge invariant one needs to
add an open Wilson line inside the trace. We will discuss such operators 
in more detail in section \ref{sec:nonlocality}.

\section{Closed Wilson loops and Wilson surfaces}
\label{sec:Wilson}
In this section we use the dual gravity description of nonlocal  
theories studied in the previous sections to compute the 
expectation values of Wilson loops for different theories.
According to the AdS/CFT 
correspondence the expectation value of the Wilson loop
of the gauge theory can be computed in the
dual string theory description by evaluating the partition function
of a string whose worldsheet is bounded by the loop \cite{{Rey:1998ik},
{Maldacena:1998im}}. 
In the supergravity approximation the dominant contribution comes from the 
minimal two dimensional surface bounded by the loop. 
The expectation value of the Wilson loop is
\be
 \left\langle W(C) \right\rangle \sim e^{-S}\;,
\ee
where S is the string action evaluated on
the minimal surface bounded by the loop $C$.

\subsection{Dipole theory}
The Wilson loop of  the dipole gauge theories living on the worldvolume 
of Dp-branes in the presence of nonzero B-field with one leg along the brane 
has been studied in \cite{Alishahiha:2002ex} using 
the supergravity solution (\ref{NEAR}).
When the distance between a quark and 
an anti-quark is much bigger than their  dipole size,
their energy is given by
\be
E\sim -\left(\frac{\gYM^2 N}{l^2}\right)^{1/(5-p)}\left(1+
c_0\Leff^2
\left(\frac{\gYM^2 N}{l^2}\right)^{2/(5-p)}+\dots\right)\;,
\label{DIDI}
\ee
where $l$ is the $Q \bar{Q}$ separation, $c_0$ is a numerical 
constant and $2\pi\Leff$ is the effective dipole vector
defined in (\ref{effdipole}).
The first term in the above expression is what we have 
in the ordinary gauge theory and the
second term can be interpreted as the dipole-dipole
interaction. Note that
in our computation leading to (\ref{DIDI}) we have kept fixed
the angular coordinates $\theta_i$ which are involved in the definition
of $\Leff$ in (\ref{effdipole}).
The angular variables are canonically dual to the R-symmetry charge
of the quark (and opposite charge of the anti-quark).
Keeping the angular variables fixed
translates to a fixed dipole electric moment for the quark and
anti-quark that is given by $2\pi\Leff$.

\subsection{Discpole theory}
Let us now compute the Wilson surface of the discpole theory 
described by the supergravity solution (\ref{M5D}).
{}From the Wilson surface the potential per unit length of two 
external straight string-like objects can be calculated.
To do this we recognize four different membrane configurations 
which we parameterize as follows
\begin{enumerate}
	\item $t=\tau,\;\;\;x_4=\sigma_1,\;\;\;\;x_1=\sigma_2\equiv x,\;\;\;\;
	u=u(x)$ at $\theta_1=\theta_0$=constant.
	
	\item $t=\tau,\;\;\;x_1=\sigma_1,\;\;\;\;x_4=\sigma_2\equiv x,\;\;\;\;
	u=u(x)$ at $\theta_1=\theta_0$=constant.
	
	\item $t=\tau,\;\;\;x_1=\sigma_1,\;\;\;\;x_2=\sigma_2\equiv x,\;\;\;\;
	u=u(x)$ at $\theta_1=\theta_0$=constant.
	
	\item $t=\tau,\;\;\;x_4=\sigma_1,\;\;\;\;x_5=\sigma_2\equiv x,\;\;\;\;
	u=u(x)$ at $\theta_1=\theta_0$=constant.
		
\end{enumerate}
\begin{figure}[h]
\begin{picture}(400,300)
\put(0,150){\begin{picture}(200,100)
\thicklines
\put(-10,10){\line(1,0){100}} %
\put(95,50){\line(1,0){75}} %
\put(18,15){$1^{st}$ object}
\put(100,55){$2^{nd}$ object}

\thinlines
\qbezier(0,10)(30,110)(40,110) %
\qbezier(80,10)(110,110)(120,110) %
\qbezier(120,110)(130,110)(160,50) %
\put(40,110){\line(1,0){80}} %

\thinlines
\put(100,10){\vector(0,1){20}} %
\put(100,10){\vector(1,0){20}} %
\put(100,10){\vector(2,1){20}} %
\put(98,35){$u^{-1}$}
\put(122,8){$x_4$}
\put(122,20){$x_1$}

\put(80,120){case (1)}
\end{picture}}

\put(200,150){\begin{picture}(200,100)
\thicklines
\put(-10,10){\line(1,0){100}} %
\put(95,50){\line(1,0){75}} %
\put(18,15){$1^{st}$ object}
\put(100,55){$2^{nd}$ object}

\thinlines
\qbezier(0,10)(30,110)(40,110) %
\qbezier(80,10)(110,110)(120,110) %
\qbezier(120,110)(130,110)(160,50) %
\put(40,110){\line(1,0){80}} %

\thinlines
\put(100,10){\vector(0,1){20}} %
\put(100,10){\vector(1,0){20}} %
\put(100,10){\vector(2,1){20}} %
\put(98,35){$u^{-1}$}
\put(122,8){$x_1$}
\put(122,20){$x_4$}
\put(80,120){case (2)}
\end{picture}}

\put(0,0){\begin{picture}(200,100)
\thicklines
\put(-10,10){\line(1,0){100}} %
\put(95,50){\line(1,0){75}} %
\put(18,15){$1^{st}$ object}
\put(100,55){$2^{nd}$ object}

\thinlines
\qbezier(0,10)(30,110)(40,110) %
\qbezier(80,10)(110,110)(120,110) %
\qbezier(120,110)(130,110)(160,50) %
\put(40,110){\line(1,0){80}} %

\thinlines
\put(100,10){\vector(0,1){20}} %
\put(100,10){\vector(1,0){20}} %
\put(100,10){\vector(2,1){20}} %
\put(98,35){$u^{-1}$}
\put(122,8){$x_1$}
\put(122,20){$x_2$}
\put(80,120){case (3)}
\end{picture}}

\put(200,0){\begin{picture}(200,100)
\thicklines
\put(-10,10){\line(1,0){100}} %
\put(95,50){\line(1,0){75}} %
\put(18,15){$1^{st}$ object}
\put(100,55){$2^{nd}$ object}

\thinlines
\qbezier(0,10)(30,110)(40,110) %
\qbezier(80,10)(110,110)(120,110) %
\qbezier(120,110)(130,110)(160,50) %
\put(40,110){\line(1,0){80}} %

\thinlines
\put(100,10){\vector(0,1){20}} %
\put(100,10){\vector(1,0){20}} %
\put(100,10){\vector(2,1){20}} %
\put(98,35){$u^{-1}$}
\put(122,8){$x_4$}
\put(122,20){$x_5$}
\put(80,120){case (4)}
\end{picture}}

\end{picture}
\caption{Four different orientations of the membrane in $AdS$.
The nonlocality is in directions $x_4,x_5.$}
\label{fig:MembCases}
\end{figure}
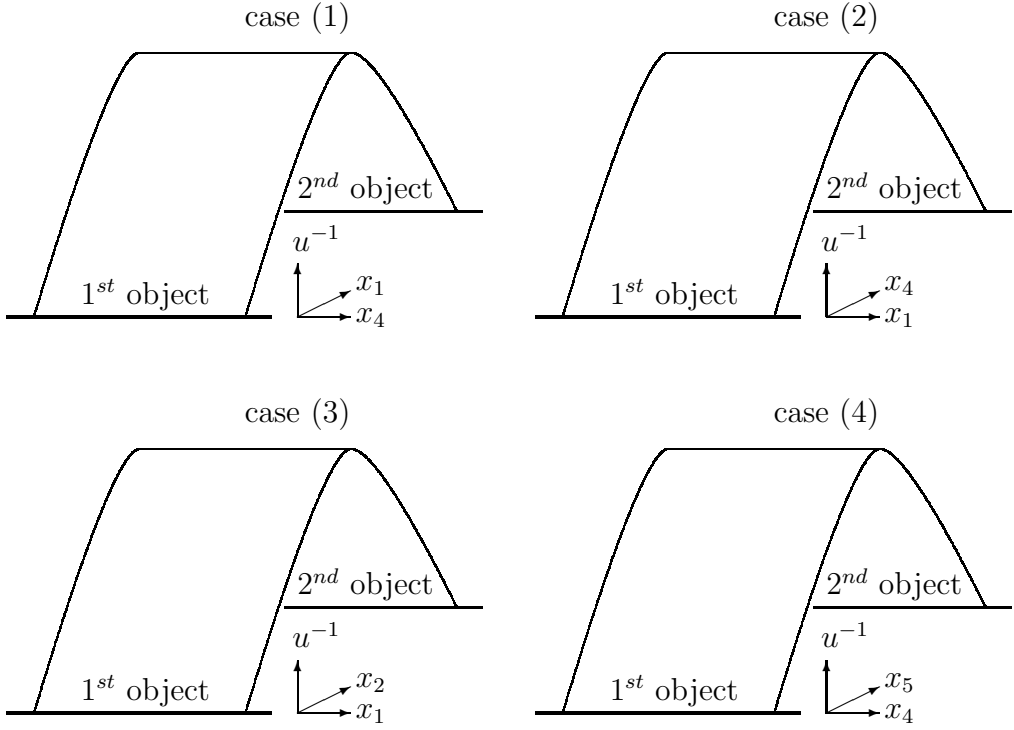

Here the membrane worldvolume is parametrized by $\tau,\sigma_1$ and
$\sigma_2$. The configurations are depicted in figure \ref{fig:MembCases}.
We are interested in the action per unit $\Delta\sigma_1\Delta\tau$ area.
We therefore consider a finite piece of the membrane given by
$0\le \sigma_1\le L'$ and $0\le \tau\le T$. In the supergravity approximation
the action should be proportional to $L' T$.
For the case (1) the membrane action
\be
S={1\over (2\pi)^2l_p^3}\int d\sigma^3\sqrt{-\det(G_{\mu\nu}\partial_aX^{\mu}
\partial_bX^{\nu})},
\ee
reads
\be
S={TL'\over (2\pi)^2}\int dx\sqrt{{u^3\over R^3}+(\partial_xu)^2}
\ee
which is exactly the same as that considered in \cite{Maldacena:1998im}. 
Therefore one finds
\be
{E\over L'}\sim -{\pi N\over l^2}
\ee
where $l$ is the distance between two external string-like objects of the
theory.
This means that when the external objects are perpendicular to the
$4-5$ plane (the two nonlocality directions) and the plane that passes
through the object intersects the 4-5 plane in a line (the $4^{th}$ direction)
as represented by
membrane configuration in case (1) the force between them is
is not sensitive to the dipole deformation effect.

For the other cases, the membrane actions read

\begin{itemize}
	\item Case 2 
\be
S={TL'\over (2\pi)^2}\int dx\sqrt{{u^3\over R^3}+(1+u^2\Leff^2)
(\partial_xu)^2}\;.
\ee

	\item Case 3 
\be
S={TL'\over (2\pi)^2}\int dx\sqrt{(1+u^2\Leff^2)\left({u^3\over R^3}+
(\partial_xu)^2\right)}\;.
\ee

\item Case 4 
\be
S={TL'\over (2\pi)^2}\int dx\sqrt{{u^3/R^3\over 1+u^2\Leff^2}+
(\partial_xu)^2}\;.
\ee

\end{itemize}
The above formulas use the effective discpole $2\pi\Leff$ defined in
(\ref{LeffDP}). 
Using the same method as \cite{Maldacena:1998im} the energy as the function of 
$l$ for all cases is given by
\be\label{QQAllCases}
{E\over L'}\sim -{\pi N\over l^2}\left(1+c_0^i\Leff\;
{(\pi N)^2\over l^4}\cdots\right)\;,
\ee
where $c_0^i,\;i=2,3,4$ are some numerical constants which, of course,
depend on 
the case number ($2,3$ or $4$)
that we are considering. The first term in the above expression is the
normal interaction between external object in the (2,0) theory. The
second term can be interpreted as the discpole interaction between the 
objects. So the corresponding object represented by the membrane 
configuration given by cases 2-4 in addition to the normal ${1\over l^2}$
interaction have discpole-discpole interaction as well. Therefore the
corresponding states must have discpole moment. 

\subsection{Lightlike dipole theory}
Now we would like to study the potential of the $Q \bar{Q}$ 
system for the lightlike dipole theory using the corresponding
supergravity solution (\ref{lltwmet}).
We parameterize the string configuration by $x^+ =\tau, 
x_1=\sigma=x, u=u(x)$ for fixed $x^-$. We also keep fixed the
angular coordinates which  are involved in the definition of
$L_{{\rm eff}}$. In this parameterization, using the supergravity 
solution (\ref{lltwmet}), the string action 
\be
S= \frac{1}{2 \pi l_s^2} 
\int d \tau  d \sigma \sqrt{-\det \left( G_{\mu \nu} 
\partial_i X^{\mu} \partial_j X^{\nu} \right)}\;. 
\label{Wilson}
\ee
reads
\be
S= \frac{T}{2\pi}{\mueff\over 2} \int dx\;u \sqrt{{(u/R)^{7-p}}
+(\partial_xu)^2}\;.
\label{WILAC}
\ee
Here we used the lightlike theory's $\mueff$
defined in equation (\ref{mueffdef}).
{}From the form of the action we find that 
despite the fact that the theory is non-local,
the end-points of the string can be fixed at 
large $u$. We note that in the noncommutative gauge theory
where we have a non-zero B field with both legs along
the brane worldvolume we have a problem fixing the end-points 
\cite{Maldacena:1999mh}, though we could fix it using a moving 
frame \cite{Alishahiha:1999ci}.

The action (\ref{WILAC}) is minimized when
\be 
u\frac{(u/R)^{7-p}}{\sqrt{{(u/R)^{7-p}}
+(\partial_xu)^2}}=u_0({u_0\over R})^{(7-p)/2}\;,
\ee
where $u_0$ is the point where $\partial_xu|_{u_0}=0$. 
This equation can be solved for $\partial_ux$, and from that
the $Q \bar{Q}$ separation is found to be
\be
{l\over 2}\defineas x(u\rightarrow \infty)= \frac{R^{(7-p)/2}}
{u_0^{(5-p)/2}}\int_{1}^{\infty}\frac{dy}{y^{(7-p)/2}
\sqrt{y^{9-p}-1}}\;.
\ee
Using (\ref{WILAC}) we can calculate the
energy of the $Q \bar{Q}$ system as follows
\be
E={\mueff\over 4\pi}u_0^2\bigg{[}\int_1^{\infty}dy
\bigg{(}\frac{y^{11-p}}{\sqrt{y^{9-p}-1}}-y\bigg{)}-{1\over 2}
\bigg{]}
\ee
Here we subtracted the infinity coming from mass of the 
W-boson which corresponds to string stretching all 
the way to $u=\infty$. 

Therefore the energy as a function of $l$ is obtained
\be\label{QQELL}
E\sim -{\mueff\over 4\pi}\;\left(\frac{g_{YM}^2N}{l^2}\right)^{2/(5-p)}.
\ee
This means that the objects' interaction with each other is only due to 
the lightlike dipoles they are carrying.

\subsection{Lightlike discpole theory}
Now we would like to study the Wilson surface in the discpole theory
described by the supergravity solution (\ref{LLDISC}). Consider an open 
membrane solution in the background (\ref{LLDISC}) parameterizing as 
following
\be
x^+=\tau,\;\;\;\;x_1=\sigma_1,\;\;\;\;x_2=\sigma_2\equiv x,\;\;\;\;u=u(x),
\;\;\;\;x^-={\rm constant}\;.
\ee 
Using the lightlike discpole theory's $\mueff$
defined after equation (\ref{LLDISC}) we find that
the membrane action for this configuration is given by
\be
S={TL'\over (2\pi)^2}{\mueff\over 2}\int dx\;u\; 
\sqrt{{u^3\over R^3}+(\partial_xu)^2}\;,
\ee
which is minimized when
\be
\frac{u^4/R^3}{\sqrt{{u^3\over R^3}+(\partial_xu)^2}}
={\rm constant}={u_0^{5/2}\over R^{3/2}}\;.
\ee
Here $u_0$ is the point where $\partial_xu|_{u_0}=0$.
This equation can be solved for $\partial_x u$ and thereby
the separation between two external objects is given by
\be
{l\over 2}\defineas x(u\rightarrow \infty)=
{R^{3/2}\over u_0^{1/2}}\int_{1}^{\infty}
\frac{dy}{y^{3/2}\sqrt{y^5-1}}\;.
\ee
Performing the integral one finds
\be
l= \frac{2\Gamma(\frac{3}{5})}{5\Gamma(\frac{1}{10})}
\sqrt{\frac{\pi R^3}{u_0}} = 
(0.55\dots)\;{2R^{3/2}\over u_0^{1/2}}\;.
\label{ll}
\ee 
Using the expression for the membrane action we can calculate the 
energy of the system as
\be 
{E\over L'}={\mueff\over 2 (2\pi)^2}\;u_0^2
\bigg{[}\int_{1}^{\infty}dy\left(\frac{y^{7/2}}
{\sqrt{y^5-1}}-y\right)-{1\over 2}\bigg{]}\;.
\ee
Performing the integral one finds
\be
{E\over L'}
= -(0.14\dots)\;{\mueff\over 2 (2\pi)^2}\;u_0^2
\label{ee}
\ee
{}From (\ref{ll}) and (\ref{ee}) we get
\be\label{QQELLDP}
{E\over L'}=-(0.1\dots)\;{\mueff\over 4}\;{N^2\over l^4}\;.
\ee
Similarly to the lightlike dipole theory, this shows that the external
objects of the lightlike discpole deformation of the (2,0) theory 
interact with each other only because of the lightlike discpole moment 
they are carrying. We note, however, that this is only the effect of
the infinite boost. In the undeformed (2,0) theory the 
string-like objects interact because of their charges,
while in the spacelike discpole deformation of the (2,0) theory the objects'
interaction has contributions
both from the charges and from the discpole moments that they are carrying.

\section{Nonlocality in the large $N$ limit}
\label{sec:nonlocality}
\begin{figure}[t]
\begin{picture}(400,100)

\thicklines
\put(10,20){\vector(1,0){250}} 
\put(10,50){\vector(1,0){250}} 

\put(20,20){\vector(2,1){60}} 

\thinlines
\multiput(190,20)(12,6){5}{\line(2,1){10}} 
\multiput(130,20)(12,6){5}{\line(2,1){10}} 
\multiput(70,20)(12,6){5}{\line(2,1){10}} 

\thicklines
\put(250,50){\vector(0,1){10}} 
\qbezier(250,60)(250,70)(270,70)
\qbezier(270,70)(280,70)(280,40)
\qbezier(280,40)(280,0)(230,0)
\qbezier(230,0)(190,0)(190,10)
\put(190,10){\vector(0,1){10}} 

\put(70,15){\line(0,1){5}}
\put(67,5){$x$}
\put(130,15){\line(0,1){5}}
\put(127,5){$x+2\pi L$}
\put(40,35){$\fb$}

\end{picture}
\caption{\it The boundary of the nonlocal dipole theory is a fibration
of the $x$ (horizontal) direction over an internal $\fb$ (vertical)
direction with identification of points separated by $2\pi L$.}
\label{fig:InternalDir}
\end{figure}
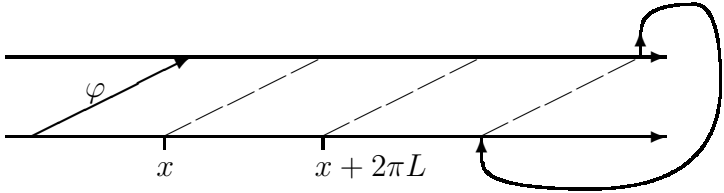

In this section we will examine how the nonlocality of the field theories
is manifested in the boundary of the supergravity duals.
In the case of the dipole theories we will show that the boundary of
the supergravity dual is better viewed as
a fibration of $\R^{3,1}$ over an internal space $M_5$
rather than simply $\R^{3,1}\times S^5$.
$M_5$ is obtain by applying T-duality along 3 internal directions in $S^5$.
In general the fibration
has a singular locus but away from that locus the fibration is
smooth and has a structure group that is generated by a finite number
of translations in $\R^{3,1}$. The translation vectors can be identified
with the dipole-vectors of the theory. It means that pairs of points
in $\R^{3,1}$ that are separated by an integral product
of a dipole-vector are connected by a path through $M_5$.
This is depicted in figure \ref{fig:InternalDir} where $\fb$ is 
a compact direction in $M_5$.
This picture was demonstrated in \cite{Bergman:2001rw}
for a special case and 
we will show this in subsection \ref{subsec:NLDT} for the general
dipole theory.

\begin{figure}[t]
\begin{picture}(250,110)
\thicklines
\qbezier(30,20)(20,50)(30,80) %
\qbezier(30,80)(40,110)(70,110) %
\qbezier(70,110)(100,110)(110,80) %
\qbezier(110,80)(120,50)(110,20) %

\put(30,20){\circle*{4}}
\put(14,9){$x-\pi L$}
\put(110,20){\circle*{4}}
\put(94,9){$x+(2n-1)\pi L$}
\put(50,95){$C$}
\put(70,110){\vector(1,0){0}}
\put(30,80){\vector(1,3){0}}
\put(110,80){\vector(1,-3){0}}

\thinlines
\multiput(31,20)(8,0){10}{\line(1,0){6}}
\put(110,20){\vector(1,0){120}}
\put(232,17){$x$}
\end{picture}
\caption{\it Open Wilson line with R-charge.
The opening is a vector in the $x$ direction with
length proportional to the R-charge.}
\label{fig:OpenWL}
\end{figure}
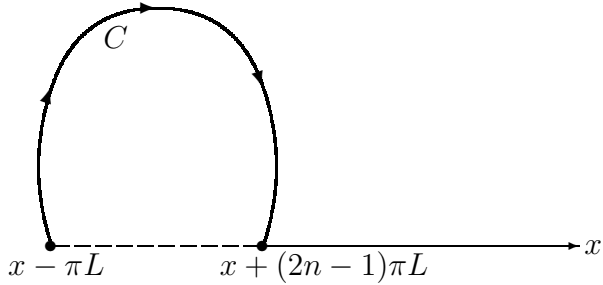

Another aspect of the nonlocality of the dipole theories is
the existence of {\it open Wilson lines} depicted in figure \ref{fig:OpenWL}.
Consider a field $\Phi$ with dipole vector $2\pi\vL$.
Let $C$ be an open path from $\vec{x} +(2n-1)\pi\vL$ 
to $\vec{x}-\pi\vL$ where $n\in\Z_{+}$ is a positive integer.
We can define the gauge invariant operator
\be\label{OpenWL}
W(C)=\tr{P e^{i\int_C A}
\Phi(\vec{x})\Phi(\vec{x}+2\pi\vL)\cdots\Phi(\vec{x}+2\pi n\vL)}.
\ee
Since $2\pi\vL$ is proportional to the R-charge
a generic gauge invariant operator with R-charge 
must have an open Wilson line with the opening proportional to
the R-charge.

\begin{figure}[t]
\begin{picture}(230,140)

\thicklines
\put(160,10){\line(-2,1){40}} %
\put(120,30){\line(2,1){40}} %
\put(160,50){\line(2,-1){40}} %
\put(200,30){\line(-2,-1){40}} %
\qbezier(200,30)(200,50)(210,70) 
\qbezier(120,30)(120,50)(110,70) 
\qbezier(110,70)(100,90)(110,110) 
\qbezier(210,70)(220,90)(210,110) 
\qbezier(210,110)(200,130)(180,140) 
\qbezier(110,110)(120,130)(140,140) 
\qbezier(180,140)(160,150)(140,140) 

\thinlines


\put(155,115){\begin{picture}(40,40)
  \qbezier(5,20)(-5,15)(-10,5) 
  \qbezier(10,15)(0,10)(-5,0) 
  \qbezier(-5,20)(5,15)(10,5) 
  \qbezier(-10,15)(0,10)(5,0) 
  \end{picture}}

\put(125,75){\begin{picture}(40,40)
  \qbezier(0,20)(-5,10)(-5,5) 
  \qbezier(5,15)(0,5)(0,0) 
  \qbezier(-5,20)(0,10)(10,5) 
  \qbezier(-10,15)(-5,5)(5,0) 
  \end{picture}}

\put(205,70){\begin{picture}(40,40)
  \qbezier(-15,20)(-10,10)(-10,5) 
  \qbezier(-10,25)(-5,15)(-5,10) 
  \qbezier(0,20)(-5,10)(-15,5) 
  \qbezier(-5,25)(-10,15)(-20,10) 
  \end{picture}}

\qbezier(160,50)(160,70)(150,80) %

\thinlines
\multiput(160,50)(4,-2){10}{\line(-2,-1){40}} 
\multiput(160,50)(-4,-2){10}{\line(2,-1){40}} 

\put(120,30){\vector(-2,1){20}}
\put(95,45){$x_4$}
\put(200,30){\vector(2,1){20}}
\put(222,40){$x_5$}

\end{picture}
\caption{\it Open Wilson surface with R-charge.
The boundary traces out a closed curve in the $x_4,x_5$ plane
that bounds an area proportional to the R-charge.
The opening (depicted as a cross hatched square) could be of any shape.}
\label{fig:OpenWSurface}
\end{figure}
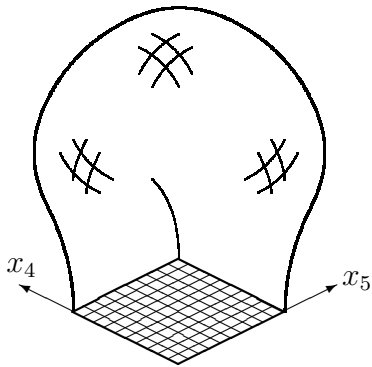

Moving on to the generalized nonlocal theories,
we conjecture that there are operators with R-charge
that correspond to open manifolds.
For example, consider the discpole theory which is a nonlocal
deformation of the $(2,0)$ theory.
The $(2,0)$ theory is believed to have {\it Wilson surface} operators,
that is, operators that correspond to closed surfaces.
Consider the nonlocal deformation that assigns an area in the $x_4,x_5$
plane to R-charge. Such a theory, we conjecture, will have operators
that correspond to open surfaces.  A typical open-Wilson-surface operator
has a  boundary that lies in a single plane
parallel to the $x_4,x_5$ directions.
Such a surface is depicted in figure \ref{fig:OpenWSurface}.
By analogy with the open Wilson lines of dipole theories,
the area of the opening will be proportional to the R-charge of
the operator.
Presumably, such operators can have a disconnected boundary $\bigcup_i D_i$
where each $D_i$ is a curve on a plane of constant $x_0, x_1, x_2, x_3$
and the sum
of the (signed) areas bounded by all the $D_i$'s is proportional to the
R-charge.

\subsection{Nonlocality in dipole theories}
\label{subsec:NLDT}
We parameterize the matrix of dipole vectors $\hM$, defined in 
subsection \ref{subsec:funtw}, as the matrix
\be\label{hMgeneric}
\hM=
\left(\begin{array}{llllll}
0 & L_1 & 0 & 0 & 0 & 0 \\
-L_1 & 0 & 0 & 0 & 0 & 0 \\
0 & 0 & 0 & L_2 & 0 & 0 \\
0 & 0 & -L_2 & 0 & 0 & 0 \\
0 & 0 & 0 & 0 & 0 & L_3 \\
0 & 0 & 0 & 0 & -L_3 & 0 \\
\end{array}\right)
\ee
The parameters $2\pi L_1, 2\pi L_2, 2\pi L_3$ are the dipole lengths of the 
scalar fields. The dipole lengths of the fermionic fields are
\bear
\lam_1 &\defineas& \pi(L_1 -L_2 -L_3),\quad
\lam_2 \defineas \pi(L_2 -L_1 -L_3),\quad
\lam_3 \defineas \pi(L_3 -L_1 -L_2),
\nn\\
\lam_4 &\defineas& \pi (L_1 +L_2 +L_3) = -(\lam_1 + \lam_2 + \lam_3).
\nn
\eear
To see how all these distances appear as the nonlocality parameters
we start with 
the metric and NSNS $B$-field in equation (\ref{NEAR}) for $p=3$.
We can parameterize $S^5$ as a $T^3$ fibration over $S^2$,
as in \cite{Alishahiha:2002ex},
$$
\nv = (
\yb_1\sin\fb_1,\,
\yb_1\cos\fb_1,\,
\yb_2\sin\fb_2,\,
\yb_2\cos\fb_2,\,
\yb_3\sin\fb_3,\,
\yb_3\cos\fb_3),
$$
with $\yb_1^2 + \yb_2^2 +\yb_3^2 = 1,$ where
$(\yb_1, \yb_2, \yb_3)$ parameterize the base $S^2$.
The metric and B-field (\ref{NEAR}) can now be written as
\bear
 {\a'}^{-1} ds^{2} &=& \frac{u^2}{R^2}
     \left(dt^2 -dx_{1}^{2} - dx_{2}^{2} 
-\frac{dx_3^2}{1+u^2\sum_{i=1}^3 L_i^2 \yb_i^2}\right)
-R^2 \left(d\Omega_2^2 +\frac{du^2}{u^2}\right)
\nn\\ &&
-R^2 \left\lbrack\sum_{i=1}^3 \yb_i^2 d\fb_i^2 
-\frac{u^2}{1+u^2\sum_{i=1}^3 L_i^2 \yb_i^2}
  \left(\sum_{i=1}^3 L_i \yb_i^2 d\fb_i\right)^2\right\rbrack,
\nn\\
B &=& 
\frac{u^2}{1+u^2\sum_{i=1}^3 L_i^2 \yb_i^2}
  \sum_{i=1}^3 L_i \yb_i^2 d\fb_i\wdg dx_3,\nn\\
e^{2\phi} &=& 
\frac{g_s^2}{1+u^2\sum_{i=1}^3 L_i^2 \yb_i^2}.
\label{dsGenFib}
\eear
Here $d\Omega_2^2$ is the metric on the base $S^2$.
For simplicity let us work with lightlike dipole vectors.
The background (\ref{dsGenFib}) is replaced by:
\bear
 {\a'}^{-1}ds^{2} &=& \frac{u^2}{R^2}
     \left\lbrack dx^{+}dx^{-} -dx_{1}^{2} - dx_{2}^{2} 
-\frac{u^2}{4}\left(\sum_{i=1}^3 L_i^2 \yb_i^2\right)(dx^{+})^2
\right\rbrack
\nn\\ &&
-R^2 \left(\frac{du^2}{u^2}
+d\Omega_2^2 +\sum_{i=1}^3 \yb_i^2 d\fb_i^2 \right),
\nn\\
B &=& 
u^2 \sum_{i=1}^3 L_i \yb_i^2 d\fb_i\wdg dx^{+},\nn\\
e^{2\phi} &=& g_s^2.
\label{dsLLFib}
\eear
Next we perform T-duality on the fiber $T^3$.
The metric (\ref{dsLLFib}) has an isometry $\fb_i\rightarrow\fb_i+\epsilon_i$
but for each $i$ the isometry $\fb_i\rightarrow\fb_i+2\pi$ acts as
$(-)^F$. It multiplies the fermion fields by
$(-1)$ because the isometry has fixed points near which
it acts as a $2\pi$ rotation of the tangent plane.
Because of the $(-)^F$ T-duality acts somewhat differently than
in the usual case. It takes us from type-IIB to type 0A
and the T-dual $T^3$ is smaller by a factor of $2$.
More precisely, let $P_1, P_2, P_3\in\Z$ be the momentum
generators along the circles parameterized by $\fb_1, \fb_2, \fb_3$.
The circles have radii $R \yb_i$.
For the purpose of obtaining the exact T-dual background
we can double the radii to make them $2R \yb_i$
and then restrict to states for which
\be\label{FPPP}
(-)^F e^{\pi i P_1}
=(-)^F e^{\pi i P_2}
=(-)^F e^{\pi i P_3} = 1.
\ee
after T-duality we get type-IIA with the background
\bear
{\a'}^{-1}ds^{2} &=& \frac{u^2}{R^2}
     \left\lbrack
dx^{+}(dx^{-} -\frac{1}{2}\sum_i L_i d\tfb_i) -dx_{1}^{2} - dx_{2}^{2}
\right\rbrack
\nn\\ &&
-R^2 \left(\frac{du^2}{u^2}
+d\Omega_2^2\right)
-\frac{1}{4R^2}\sum_{i=1}^3 \frac{d\tfb_i^2}{\yb_i^2}
\nn\\ &&
\label{xFibPhi}
\eear
Here $(\tfb_1,\tfb_2,\tfb_3)$ are coordinates on the dual $T^3$
with the identifications $\tfb_i\sim \tfb_i+2\pi$,
but we have to augment the background with the T-dual of
the projection (\ref{FPPP}):
\be\label{FWWW}
(-)^F e^{\pi i W_1}
=(-)^F e^{\pi i W_2}
=(-)^F e^{\pi i W_3} = 1.
\ee
Here $W_1, W_2, W_3$ are string winding numbers.
This means that permissible string states have winding numbers
that satisfy
$$
W_i-W_j\in 2\Z,\qquad W_i\in\Z,\qquad i,j=1,2,3,
$$
and the state is fermionic if all $W_i$ are odd
and bosonic if all $W_i$ are even.
{}From this it follows that the compactification is
actually type-0A and the $T^3$ fibers are the spaces
parameterized by $(\tfb_1,\tfb_2,\tfb_3)$ with the identifications
$$
(\tfb_1,\tfb_2,\tfb_3)\sim
(\tfb_1+2\pi n_1,\tfb_2+2\pi n_2,\tfb_3+2\pi n_3),\qquad
n_i\in\Z,\qquad n_i-n_j\in 2\Z.
$$
We can define
$$
\xi^{-}\defineas
  x^{-} -\frac{1}{2}\sum L_i \tfb_i,\qquad \xi^{+}\defineas x^{+}.
$$
The metric (\ref{xFibPhi}) is then simply
\bear
{\a'}^{-1}ds^{2} &=& \frac{u^2}{R^2}
\left\lbrack
d\xi^{-}d\xi^{+} -dx_{1}^{2} - dx_{2}^{2}
\right\rbrack
-R^2 \left(\frac{du^2}{u^2}
+d\Omega_2^2\right)
-\frac{1}{4R^2}\sum_{i=1}^3 \frac{d\tfb_i^2}{\yb_i^2}.
\nn\\ &&
\label{FibPhixi}
\eear
It describes the $\xi^{-}$ direction as fibered
over the $T^3$ in such a way that $\tfb_i\rightarrow\tfb_i+2\pi n_i$ 
is accompanied by $\xi^{-}\rightarrow \xi^{-}-\pi\sum_i n_i L_i,$
with $n_1, n_2, n_3$ all odd or all even.
The full identification should therefore be
\bear
(\tfb_1,\tfb_2,\tfb_3,\xi^{+}) &\sim&
(\tfb_1+2\pi n_1,\tfb_2+2\pi n_2,\tfb_3+2\pi n_3,
\xi^{+}-\pi\sum_i n_i L_i),
\nn\\
&& n_i\in\Z,\qquad n_i-n_j\in 2\Z.\label{identphxi}
\eear
It is therefore obvious from the structure of spacetime in the dual
supergravity that there are nonlocal interactions between fields at
distance $2\pi L_i$ ($i=1\dots 3$) and at distance $\lam_a$ ($a=1\dots 4$).

%

For completeness we will also present the T-dual background to
the general background (\ref{dsGenFib}). It is given by type-0A
with metric
\bear
 {\a'}^{-1}ds^{2} &=& \frac{u^2}{R^2}
     \left(dx_0^2 -dx_{1}^{2} - dx_{2}^{2}\right)
-R^2 \left(d\Omega_2^2 +\frac{du^2}{u^2}\right)
\nn\\ &&
-\frac{u^2}{R^2}\left(dx_3-\frac{1}{2}\sum_{i=1}^3 L_i d\tfb_i\right)^2
-\frac{1}{R^2}\sum_{i=1}^3 \frac{d\tfb_i^2}{\yb_i^2}
\nn\\
e^{2\phi} &=& 
\frac{g_s^2}{R^6\prod_{i=1}^3 \yb_i^2},\qquad\qquad
B = 0,
\label{dsGenFibTD}
\eear
and with the identifications
\be
(\tfb_1,\tfb_2,\tfb_3) \sim
(\tfb_1+2\pi n_1,\tfb_2+2\pi n_2,\tfb_3+2\pi n_3)
\qquad
n_i\in\Z,\qquad n_i-n_j\in 2\Z.\label{identph}
\ee

\subsection{Open Wilson lines in dipole theories}\label{subsec:OWLD}
\begin{figure}[t]
\begin{picture}(300,120)
\thicklines

\put(30,20){
  \begin{picture}(170,110)
  \thicklines
  \qbezier(0,0)(-10,30)(0,60) %
  \qbezier(0,60)(10,90)(40,90) %
  \qbezier(40,90)(70,90)(80,60) %
  \qbezier(80,60)(90,30)(80,0) %
  \put(0,0){\circle*{4}}
  \put(80,0){\circle*{4}}
  \put(20,75){$C_1$}
  \put(40,90){\vector(1,0){0}}
  \put(0,60){\vector(1,3){0}}
  \put(80,60){\vector(1,-3){0}}
  \put(78,0){\vector(1,0){0}}
  \thinlines
  \multiput(1,0)(8,0){10}{\line(1,0){6}}
  \put(32,4){$2\pi\vec{L}_1$}
  \end{picture}}

\put(150,50){
  \begin{picture}(170,110)
  \thicklines
  \qbezier(0,0)(-5,15)(0,30) %
  \qbezier(0,30)(5,45)(20,45) %
  \qbezier(20,45)(35,45)(40,30) %
  \qbezier(40,30)(45,15)(40,0) %
  \put(0,0){\circle*{4}}
  \put(40,0){\circle*{4}}
  \put(10,35){$C_2$}
  \put(20,45){\vector(-1,0){0}}
  \put(0,30){\vector(-1,-3){0}}
  \put(40,30){\vector(-1,3){0}}
  \put(2,0){\vector(-1,0){0}}
  \thinlines
  \multiput(1,0)(8,0){5}{\line(1,0){6}}
  \put(10,4){$2\pi\vec{L}_2$}
  \end{picture}}

\put(230,70){
  \begin{picture}(170,110)
  \thicklines
  \qbezier(0,0)(-5,-15)(0,-30) %
  \qbezier(0,-30)(5,-45)(20,-45) %
  \qbezier(20,-45)(35,-45)(40,-30) %
  \qbezier(40,-30)(45,-15)(40,0) %
  \put(0,0){\circle*{4}}
  \put(40,0){\circle*{4}}
  \put(10,-37){$C_3$}
  \put(20,-45){\vector(-1,0){0}}
  \put(0,-30){\vector(-1,3){0}}
  \put(40,-30){\vector(-1,-3){0}}
  \put(2,0){\vector(-1,0){0}}
  \thinlines
  \multiput(1,0)(8,0){5}{\line(1,0){6}}
  \put(10,4){$2\pi\vec{L}_3$}
  \end{picture}}
\end{picture}
\caption{\it Correlation function of open Wilson lines.
It can be nonzero if $\vec{L}_1 + \vec{L}_2 + \vec{L}_3 = 0.$}
\label{fig:ManyOpenWL}
\end{figure}
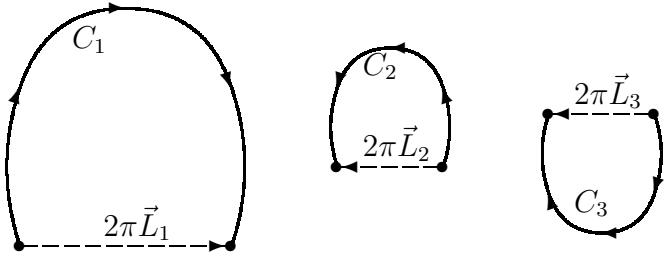

\begin{figure}[t]
\begin{picture}(400,120)

\put(10,60){\begin{picture}(200,50)
  \thinlines
  \put(0,0){\vector(1,0){250}} 
  \put(0,30){\vector(1,0){250}} 
  \put(10,0){\vector(2,1){60}} 
  \thinlines
  \multiput(60,0)(12,6){5}{\line(2,1){10}} 
  \multiput(120,0)(12,6){5}{\line(2,1){10}} 
  \put(120,30){\line(0,1){5}}
  \put(117,37){$x^3$}
  \put(180,30){\line(0,1){5}}
  \put(177,37){$x^3+2\pi L$}
  \put(30,15){$\tfb$}
  \end{picture}}

\put(130,60){\begin{picture}(70,70)
  \thicklines
  \put(0,0){\circle*{4}}
  \put(-3,3){$P'$}
  \put(60,0){\circle*{4}}
  \put(63,4){$Q$}
  \put(60,30){\circle*{4}}
  \put(63,20){$P$}
  \put(45,8){$C_\tfb$}

  \qbezier(10,-50)(30,-60)(50,-50) %
  \put(50,-50){\vector(2,1){0}}
  \qbezier(10,-50)(-10,-40)(0,0) %
  \put(10,-50){\vector(2,-1){0}}
  \qbezier(50,-50)(70,-40)(60,0)
  \put(30,-50){$C$}
  \put(60,0){\vector(0,1){30}}
  \end{picture}}

\thicklines
\end{picture}
\caption{\it In the supergravity dual, an open Wilson line actually
closes through the extra $\tfb$-dimension along the path
$C_\tfb$. The dashed lines connect points that are identified.}
\label{fig:ClosedOpenWL}
\end{figure}
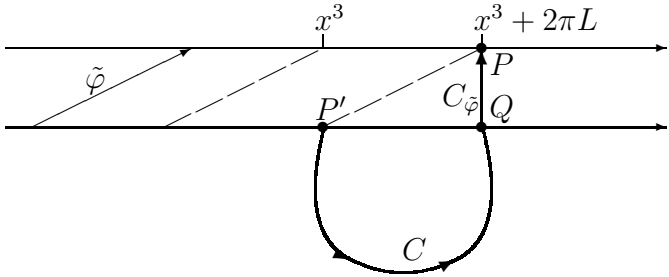

The discussion of Wilson lines is more natural in Euclidean space.
We will therefore Wick rotate the background (\ref{dsGenFib})
to obtain the Euclidean type-IIB background
\bear
 {\a'}^{-1}ds^{2} &=& \frac{u^2}{R^2}
     \left(\sum_{\u=0}^2 dx_\u^2
+\frac{dx_3^2}{1+u^2\sum_{i=1}^3 L_i^2 \yb_i^2}\right)
+R^2 \left(d\Omega_2^2 +\frac{du^2}{u^2}\right)
\nn\\ &&
+R^2 \left\lbrack\sum_{i=1}^3 \yb_i^2 d\fb_i^2 
-\frac{u^2}{1+u^2\sum_{i=1}^3 L_i^2 \yb_i^2}
  \left(\sum_{i=1}^3 L_i \yb_i^2 d\fb_i\right)^2\right\rbrack,
\nn\\
B &=& 
\frac{u^2}{1+u^2\sum_{i=1}^3 L_i^2 \yb_i^2}
  \sum_{i=1}^3 L_i \yb_i^2 d\fb_i\wdg dx_3,\nn\\
e^{2\phi} &=& 
\frac{g_s^2}{1+u^2\sum_{i=1}^3 L_i^2 \yb_i^2}.
\label{dsEuGenFib}
\eear
Similarly, we can Wick rotate the background (\ref{dsGenFibTD}) to obtain
the T-dual Euclidean type-0A background
\bear
 {\a'}^{-1}ds^{2} &=& \frac{u^2}{R^2}\sum_{\u=0}^2 dx_\u^2
+R^2 \left(d\Omega_2^2 +\frac{du^2}{u^2}\right)
\nn\\ &&
+\frac{u^2}{R^2}\left(dx_3-\frac{1}{2}\sum_{i=1}^3 L_i d\tfb_i\right)^2
+\frac{1}{R^2}\sum_{i=1}^3 \frac{d\tfb_i^2}{\yb_i^2}
\nn\\
e^{2\phi} &=& 
\frac{g_s^2}{R^6\prod_{i=1}^3 \yb_i^2},\qquad\qquad
B = 0.
\label{dsEuGenFibTD}
\eear
R-charge is conserved in dipole theories.
Unlike a mass deformation of SYM which can classically preserve
some R-symmetry but quantum mechanically instanton effects break
the classical symmetry, the classical R-symmetry that is preserved
by a dipole deformation cannot be broken by instantons.
This is because the remaining R-symmetry is a geometrical rotation
symmetry in the string theory realization.
With the generic dipole vectors given in (\ref{hMgeneric}) the
unbroken symmetry is $U(1)^3$.
Consider now a correlation function 
$\langle W(C_1) W(C_2)\cdots W(C_r)\rangle$ of open Wilson lines
$C_1, C_2,\dots, C_r$ as in (\ref{OpenWL}).
Let the openings of the Wilson lines be given by the vectors
$2\pi\vL_1, 2\pi\vL_2,\cdots, 2\pi\vL_r$. 
(See figure \ref{fig:ManyOpenWL} for an example.)
Because of R-charge conservation we need to have 
$$
0 = \vL_1 + \vL_2 + \cdots + \vL_r.
$$
In the supergravity dual (\ref{dsEuGenFibTD}) each open Wilson line corresponds
to a closed curve on the boundary. 
It is constructed from the open curve by closing it through
the extra $\tfb_1,\tfb_2,\tfb_3$ directions.
Two points $P,Q$ whose $x^3$ coordinates differ by $2\pi L$ and whose
remaining coordinates, including $\tfb_1,\tfb_2,\tfb_3,$ are identical
can be connected by a microscopic path through the
$\tfb_1,\tfb_2,\tfb_3$ dimensions (see Figure \ref{fig:ClosedOpenWL}).
Using the metric (\ref{dsEuGenFibTD})
we see that the distance 
between the points $P$ and $Q$ through the $\tfb$-direction
along the path $C_\tfb$ in figure (\ref{fig:ClosedOpenWL})
is of the order of $\frac{1}{R}$ whereas the proper
length of a curve between $P'$ and $Q$ with fixed 
$\tfb_1,\tfb_2,\tfb_3$ is of the order of
$\frac{u}{R}\rightarrow\infty$ as $u\rightarrow \infty$.
Thus, the shortcut $PQ$ through the internal $\tfb$ directions
has negligible length compared to the rest of the Wilson line.

According to the prescription of \cite{Maldacena:1998im} we have
to complete the closed Wilson loop to a surface into the bulk.
In principle, one has to perform a path integral over all worldsheets
in the bulk with the given boundary conditions but in the case
of the local $\SUSY{4}$ SYM theory, and for generic, sufficiently smooth
Wilson loops the path integral is dominated
by a single classical configuration as in \cite{Maldacena:1998im}.
This was also the case in \cite{Alishahiha:2002ex}
and in section \ref{sec:Wilson}
where we studied operators without R-charge.

What would be the analogous prescription for calculating Open Wilson lines in
the large $N$ limit of the nonlocal dipole theory?
Suppose we are given an open path $C$ on $\R^4$,
as the one from $P'$ to $Q$ in figure \ref{fig:ClosedOpenWL},
with the opening proportional to one of the dipole vectors $2\pi\vL_i$.
we can close the path $C$ to make it a closed loop by connecting
the two endpoints with a straight line through the $\tfb$-direction.
Let us denote by $C_\tfb$ the extra path that closes the loop.
Now that we have a closed loop we can do the path integral
over worldsheets in the bulk whose boundary is the closed loop.
However, since the metric in the $\tfb$-direction does not
have the prefactor $\frac{1}{u^2}$ [see (\ref{dsEuGenFibTD})]
we cannot treat $C_\tfb$ as classical.
We have to take into account the fluctuations of $C_\tfb$,
but only in the $\tfb_1, \tfb_2,\tfb_3$ directions
and not in the $x_0,\dots,x_3, \yb_1, \yb_2, \yb_3$ directions.
This is because
an arbitrarily small deformation of $C_\tfb$ in the
$x_0,\dots,x_3$ directions will change the proper length
of the path, using the metric (\ref{dsEuGenFibTD}), by terms of
order $\frac{u}{R}\rightarrow\infty$ on the boundary.
This is just as well, since if $C_\tfb$ could fluctuate into
the $\R^{4}$ directions the expectation value of the Wilson
line would not have depended on the path $C$ which would be absurd!
Also, the metric components in the directions of  $\yb_1,\yb_2,\yb_3$
are of the order of $R$ and are assumed large.

Note also that the closed loop $C\bigcup C_\tfb$
has nonzero winding number around
some of the $\tfb$-directions. This is in accord with the
equivalence between winding number and R-symmetry charge.
Thus, we have a well-defined prescription for the 
correlation function of open Wilson lines as in figure \ref{fig:ManyOpenWL}.
To calculate the correlation function
in the large $N$ limit we invoke a Born-Oppenheimer approximation
where we first fix the $(x_0,\dots,x_3, \yb_1,\dots,\yb_3, u)$
coordinates of the worldsheet with the appropriate boundary conditions.
We then treat $\tfb_1,\tfb_2,\tfb_3$ quantum mechanically 
and find the quantum partition function of these fields on the string
worldsheet. We can take Dirichlet boundary conditions for
$\tfb_1,\tfb_2,\tfb_3$ so as to fix the winding numbers in these
directions on the boundary.

\begin{figure}[h]
\begin{picture}(400,150)

  \put(0,0){\begin{picture}(200,50)
    \multiput(20,110)(150,0){2}{\begin{picture}(10,10)
      \thinlines
      \multiput(0,0)(3,0){2}{\qbezier(0,0)(2,4)(6,6)}
      \multiput(0,2)(2,3){2}{\line(1,0){6}}
    \end{picture}}


    \thicklines
    \multiput(0,40)(180,0){2}{\qbezier(0,0)(10,-20)(30,-30)} %
    \multiput(0,100)(180,0){2}{\qbezier(0,0)(10,20)(30,30)} %
    \put(0,40){\vector(-1,2){0}}
    \put(30,130){\vector(2,1){0}}
    \put(180,100){\vector(-1,-2){0}}
    \put(210,10){\vector(2,-1){0}}


    \thicklines
    \put(210,10){\qbezier(0,0)(20,0)(30,30)} %
    \put(210,130){\qbezier(0,0)(20,0)(30,-30)} %

    \multiput(60,40)(180,0){2}{\qbezier(0,0)(10,30)(0,60)} %
    \put(240,40){\vector(1,3){0}}
    \put(240,100){\vector(-1,3){0}}
    \put(60,40){\vector(-1,-3){0}}

    \put(0,100){\line(1,0){180}} %
    \put(30,130){\line(1,0){180}} %
    \put(0,40){\line(1,0){180}} %
    \put(30,10){\line(1,0){180}} %

    \put(0,40){\begin{picture}(180,60)
      \thinlines
      \multiput(3,0)(6,0){30}{\begin{picture}(5,60)
         \multiput(0,2)(0,10){6}{\line(0,1){6}}
      \end{picture}}
    \end{picture}}

    \put(5,125){$C_1$}
    \put(240,105){$C_2$}

  \end{picture}}
\end{picture}
\caption{The open worldsheet connecting two open 
Wilson lines, $C_1$ and $C_2$, can be closed with an extra piece
of microscopic action (depicted by the dashed pattern).}
\label{fig:OpenStrings}
\end{figure}
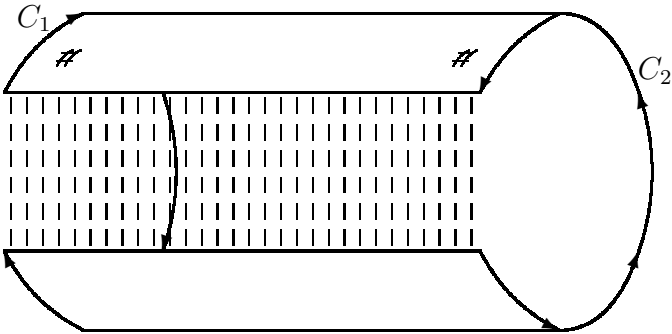

To leading order in $R$ we can ignore the coordinates
$\tfb_1,\tfb_2,\tfb_3$ and consider only the coordinates 
$x^0,\dots,x^3, u, \yb_1,\dots,\yb_3$ that describe $AdS_5\times S^2$
(see figure \ref{fig:OpenStrings}).
The prescription for calculating
the correlation function
$$
\langle W(C_1) W(C_2)\cdots W(C_r)\rangle
$$
of open Wilson lines
$C_1, C_2,\dots, C_r$ is then as follows.

Consider all possible surfaces $S\subset AdS_5\times S^2$
such that the intersection of the boundary $u=\infty$ of $AdS_5$
and $S$ is $C_1\cup\cdots\cup C_r$. $S$ is also allowed to have
a boundary $\partial S$ in the bulk but it is restricted in
the following way.
For every $0 < u_0 < \infty$
let $K_{u_0}$ be the submanifold $u=u_0$ in $AdS_5\times S^2$.
It has the geometry of $\R^4\times S^2$. It intersects
the boundary of $S$ on a collection of open paths
$$
K_{u_0}\cap \partial S = C_1^{(u_0)}\cup C_2^{(u_0)}\cup\cdots\cup
C_{m(u_0)}^{(u_0)},
$$
where the number $m(u_0)$ of disconnected paths can depend on $u_0$
because as we vary $u_0$ paths can split or join.
The restriction on $S$ is that for every $u_0$
the endpoints of each open path $C_i^{(u_0)}$ should be separated by
an integer product of the dipole vector $\vL$.
If there are several dipole vectors the separation should
be a linear combination of the dipole vectors with integer coefficients.
We also require that for $u_0$ small enough  $m(u_0)=0$ which means
that all the paths have joined together and therefore
$K_{u_0}\cap \partial S$ is either a union of closed loops or the empty set.
We now have to find the minimum area of all allowed $S$'s.
To leading order in this approximation
$$
-\log \langle
W(C_1)\cdots W(C_r)\rangle \sim \min {\mbox{Area}}(S).
$$

\subsubsection*{Example}
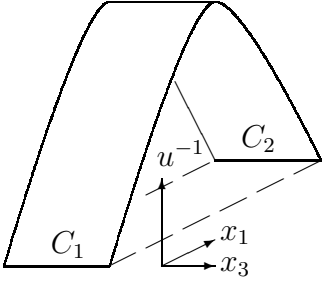
\begin{figure}[h]
\begin{picture}(230,140)
\put(0,0){\begin{picture}(90,100)
\thicklines
\put(0,10){\line(1,0){40}} %
\put(80,50){\line(1,0){40}} %
\put(18,15){$C_1$}
\put(90,55){$C_2$}

\thinlines
\qbezier(0,10)(30,110)(40,110) %
\qbezier(40,10)(70,110)(80,110) %
\qbezier(80,110)(90,110)(120,50) %
\put(40,110){\line(1,0){40}} %
\put(80,50){\line(-1,2){15}} %

\thinlines
\put(60,10){\vector(0,1){33}} %
\put(60,10){\vector(1,0){20}} %
\put(60,10){\vector(2,1){20}} %
\put(58,48){$u^{-1}$}
\put(82,8){$x_3$}
\put(82,20){$x_1$}
\multiput(40,10)(14,7){6}{\line(2,1){10}}

\put(54,37){\line(2,1){10}}
\put(68,44){\line(2,1){10}}
\end{picture}}

\end{picture}
\caption{Correlation function of two Open Wilson lines 
$\langle W(C_1) W(C_2)\rangle$. The strip is an open worldsheet in the 
$u,x_1, x_3$ space.}
\label{fig:OpenOpen}
\end{figure}

As an example we will calculate the correlation function 
$\langle W(C_1) W(C_2)\rangle$ where $C_1$ and $C_2$ are
straight segments of length $2\pi L$.
To be specific we take
\bear
\lefteqn{W(x_0, x_1, x_2, x_3) \defineas}
\nn\\ &&
\tr{
P e^{-i\int_{-\pi}^0 A_3(x_0, x_1, x_2, x_3 - t L) dt}
Z(x_0, x_1, x_2, x_3)
P e^{i\int_0^{\pi} A_3(x_0, x_1, x_2, x_3 + t L) dt}}
\nn\\ &&
\label{WPZ}
\eear
and we assume that $Z$ is an appropriate
(complex) linear combination of the 6 scalars $\Phi^I$
with a dipole vector of length $2\pi L$ in the $x_3$ direction.
We will calculate 
$$
-\log \langle W(0, -\frac{1}{2}a, 0, 0)^\dagger W(0, \frac{1}{2}a, 0, b)\rangle
$$
For this purpose we find a surface whose boundary contains the
two segments $C_1$ and $C_2$ (see figure \ref{fig:OpenOpen}).

The surface is in the space of $x_0, x_1, x_2, x_3, \yb_1, \yb_2, \yb_3$ but
the minimal area of such a surface depends only on the projection onto
the space perpendicular to the $x_3$ direction.
We will assume that $\yb_1, \yb_2, \yb_3$ are constant along $S$ and
that $x_0 = x_2 = 0$ along $S$ and we will parameterize $x_3 = y(u)$
so that $y(u_0) = 0$ is the point of $S$ with the smallest value of $u$
and that $y(\infty) = \frac{a}{2}$.
Using the metric (\ref{dsEuGenFibTD}) we see that the area of $S$ is then
\bear
\min {\mbox{Area}}(S) &=&
2 L \int_{u_0}^\infty \frac{u}{R}\sqrt{{y'}^2 + \frac{R^4}{u^4}}\, du.
\nn
\eear
This is exactly the same integral that appears in the calculation of
the quark anti-quark potential \cite{Rey:1998ik,Maldacena:1998im}.
After regularization as in \cite{Rey:1998ik,Maldacena:1998im} we 
obtain the result
$$
-\log \langle W(0, -\frac{1}{2}a, 0, 0)^\dagger W(0, \frac{1}{2}a, 0, b)\rangle
= \frac{4\pi^2\sqrt{2\gYM^2 N}}{\Gamma(\frac{1}{4})^4 a}L + \cdots
$$
where $(\cdots)$ denotes terms that are subleading when $\gYM^2 N$ is large.
We are also assuming $L\gg a$ for the following reason.
$S$ is a strip with width $L \sqrt{g_{33}} = {\a'}^{\frac{1}{2}} L u /R$.
At the points of $S$ where $u$ is smallest the proper length of the strip
is of the order of
$$
\frac{ {\a'}^{\frac{1}{2}} L u_0}{R} \sim \frac{ {\a'}^{\frac{1}{2}} L}{a},
$$
where we used the results of \cite{Rey:1998ik,Maldacena:1998im} to estimate
$u_0$. If the width of the strip is smaller than 
$l_s\defineas {\a'}^{\frac{1}{2}} $ we have to include
the contribution of the fluctuations in the $\tfb_1,\tfb_2,\tfb_3$ directions
which is a boundary effect on the strip $S$.

When  $a\gg L$ we can assume that $L$ is small.
For small $L$ the Lagrangian of the dipole theory can be written
as a small deformation of the Lagrangian of $\SUSY{4}$ SYM with gauge group
$SU(N)$ plus 6 free scalar fields and 4 free Dirac fermions.
The extra free fields come about because a generic field $Z(x)$ with
dipole vector $2\pi\vL$ transforms in the $(N,\overline{N})$ representation
of $SU(N)_{x-\pi\vL}\otimes SU(N)_{x+\pi\vL}$
where $SU(N)_x$ is the group at spacetime point $x$.
In the limit $\vL\rightarrow 0$ we find that the field $\Phi$
transforms in the reducible $(N,\overline{N})$ representation which decomposes
into the adjoint representation and a singlet.

In the limit $a\gg L$ the leading contribution to the open Wilson line $W(x)$
will be the singlet $\tr{Z}$ [see (\ref{WPZ})].
$\tr{Z}$ is a free field of conformal dimension $1$. We therefore expect
$$
-\log \langle W(0, -\frac{1}{2}a, 0, 0)^\dagger W(0, \frac{1}{2}a, 0, b)\rangle
= 2\log a + \cdots
$$

\subsection{Lightcone quantization of strings in the gravity duals
of lightlike dipole theories}
In order to extend the discussion to the generalized theories,
we would like to find the explanation for nonlocality
in terms of the original background (\ref{dsGenFib}).
A major simplification occurs in the light-like
case (\ref{dsLLFib}) when the Wilson loops are at constant $x^{-}$.

We will use the lightcone formalism for strings in $AdS_5\times S^5$
as presented in \cite{Metsaev:2000yf}-\cite{Tseytlin:2000na}.\footnote{
We wish to thank Andrei Mikhailov for pointing this out.}
Equation (21) of \cite{Tseytlin:2000na} describes the lightcone Hamiltonian
of strings in $AdS_5\times S^5$ with the metric taken as
$$
ds^2 = Y^2 dx^a dx^a + \frac{1}{Y^2}dY^K dY^K,\qquad a=0\dots 3,\qquad
K=1\dots 6.
$$
In this subsection we set $2\pi\a'=1$ for convenience. 
According to \cite{Tseytlin:2000na} the Hamiltonian density is
\bear
{\cal H} &\defineas&
\PP^{-} =
\frac{1}{2p^+}(\PP_\perp^2 + Y^4\xpr_\perp^2 +  Y^4\PP_K\PP_K +\Ypr^K\Ypr^K
\nn\\ &&
+Y^2[p^{+2}(\eta^2)^2 + 2{\rm i}p^+\eta \rho^{KN}\eta  Y_K \PP_N])
\nn\\ &&
-|Y| Y^K [\eta \rho^K (\thpr - \sqrt{2} i|Y| \eta\xpr)
+h.c.] \label{MTlcH}
\eear
Where $p^{+}$ is the constant lightcone momentum and
$Y^K$ ($K=1\dots 6$) are worldsheet fields with associated momenta
$\PP_K$. The two worldsheet fields $x_\perp$ represent 
the coordinates $x^2, x^3$ and their associated momenta are $\PP_\perp$.
$\rho^{K}$ are 6-dimensional Dirac matrices and there are two 
fermions $\th$ and $\eta$ that are also chiral $SO(6)$ spinors.
Their components are denoted by $\th_i$ and $\eta_i$ ($i=1\dots 4$)
with $\th_i^* = \th^i$ and $\eta_i^* = \eta^i$. The Dirac matrices
$\rho^K$ satisfy $\rho^K_{ij} = -\rho^K_{ji}$ and
$\rho^{KN}\defineas \rho^{[K}\rho^{\dagger N]}$.
The primes over $\thpr$, $\xpr$ and $\Ypr$ 
denote differentiation with respect to the worldsheet coordinate $\sigma$.
As is standard in lightcone gauge, the coordinate $x^{-}$ can be found
by integrating
\be\label{xminus}
\xpr^{-} = -\frac{1}{p^{+}}(\PP_\perp\xpr_\perp +\PP_K\Ypr^K)
-\frac{i}{2}(\th^i\thpr_i +\eta^i\etpr_i+\th_i\thpr^i +\eta_i\etpr^i).
\ee

It is easy to extend (\ref{MTlcH}) to the gravity dual of the lightlike
dipole theory
[(\ref{lltwmet}) with $p=3$].
When comparing (\ref{lltwmet}) to (\ref{MTlcH}) we identify
$Y^K = |Y| \nv^K$ and $|Y|^2 = u^2/R^2$.
The extra terms to add to (\ref{MTlcH}) are the contribution of a 
$\frac{u^4}{4R^2} (\nvt\cMt\cM\nv)(dx^{+})^2$ term in the metric 
and an  NSNS B-field $\frac{1}{2}u^2 d\nvt\cM\nv\wdg dx^{+}$. We obtain
the total contribution
\be
{\cal H}_1 =
\frac{p^{+}}{2}(Y^\top\cMt\cM Y + Y^\top\cM\Ypr)
+({\mbox{fermions}}).
\label{lcH1}
\ee
Here we used the gauge fixing $x^{+} = p^{+}\tau$ 
and $|Y|^2 \sqrt{g} g^{00} = 1$ (see \cite{Tseytlin:2000na})
where $(\sigma,\tau)$ are worldsheet coordinates and $g$ is the
worldsheet metric.
The expression (\ref{xminus}) for $x^{-}$ remains the same.

The extra terms (\ref{lcH1}) can be absorbed by a redefinition
\be
\vec{Y} \longrightarrow e^{i\cM\sigma}\vec{Y},\qquad
\th\longrightarrow e^{i\cM\sigma}\th,\qquad
\eta\longrightarrow e^{i\cM\sigma}\eta.
\label{SubsYThEta}
\ee
In the equation for $Y$, $\cM$ should be taken as a $6\times 6$ matrix
in the fundamental representation of $so(6)$ and in the equations
for $\th$ and $\eta$, $\cM$ should be taken as a $4\times 4$ matrix in
the spinor representation of $so(6)$.
After the substitution (\ref{SubsYThEta}) the lightcone Hamiltonian
has the same form as (\ref{MTlcH}) except that the fields $Y, \th$ and $\eta$
are no longer periodic in $\sigma$. Instead
\bear
\vec{Y}(\sigma +p^{+},\tau) &=& 
  e^{i p^{+}\cM}\vec{Y}(\sigma,\tau),\nn\\
\th(\sigma +p^{+},\tau)&=& 
  e^{i p^{+}\cM}\th(\sigma,\tau),\qquad
\eta(\sigma +p^{+},\tau)= e^{i p^{+}\cM}\eta(\sigma,\tau).
\label{Newbc}
\eear
After the redefinition (\ref{SubsYThEta}) the
expression (\ref{xminus}) for $x^{-}$ becomes:
\bear
\xpr^{-} &=& -\frac{1}{p^{+}}(\PP_\perp\xpr_\perp +\PP_K\Ypr^K)
-\frac{i}{2}(\th^i\thpr_i +\eta^i\etpr_i+\th_i\thpr^i +\eta_i\etpr^i)
\nn\\ &-&
\frac{i}{p^{+}}\PP_K \cM_{KN}Y^N
+\frac{1}{2}(\th^i {\cM_i}^j\th_j +\eta^i{\cM_i}^j\eta_j + {\mbox{c.c.}})
\label{xminustr}
\eear
Note that the second line of (\ref{xminustr}) can be written as
$\tr{\cM J^0(\sigma)}$ where 
$(J^0(\sigma),J^1(\sigma))$ are the components of the 
$so(6)$ R-symmetry worldsheet current.
We now define $\wxmin$ as
\bear
\wxmin(\sigma)  &\defineas&
\int_0^\sigma\left\lbrack
-\frac{1}{p^{+}}(\PP_\perp\xpr_\perp +\PP_K\Ypr^K)
-\frac{i}{2}(\th^i\thpr_i +\eta^i\etpr_i+\th_i\thpr^i +\eta_i\etpr^i)
\right\rbrack.
\nn
\eear
This is the same expression in terms of oscillators as (\ref{xminus})
but written in terms of the new variables 
[i.e. after the transformation (\ref{SubsYThEta})].
The new coordinate $\wxmin$ is not single valued and satisfies
\be\label{wxminbc}
\wxmin(\sigma+ p^{+}) - \wxmin(\sigma) = \tr{\cM \Rchg}
\ee
where 
$$
\Rchg^a \defineas
\int_0^{p^{+}}
\left\lbrack
-\frac{i}{p^{+}}\PP^K \tau^a_{KN}Y^N
+\frac{1}{2}(\th^i {{\tau^a}_i}^j\th_j +\eta^i{{\tau^a}_i}^j\eta_j +
{\mbox{c.c.}})
\right\rbrack,\qquad a=1\dots 15
$$
is the $so(6)$ R-symmetry charge.
[Here $\tau^a$ represents a generator of $so(6)$, $\tau^a_{KN}$
is the $6\times 6$ matrix of the generator in the representation $6$ of $so(6)$
and ${{\tau^a}_i}^j$ is its $4\times 4$ matrix in the representation $4$
of $so(6)$.]

The RHS of (\ref{wxminbc}) is exactly the expected dipole length of the
corresponding open Wilson operator.
The prescription for calculating correlation functions
of open Wilson lines in the lightlike dipole theory is as follows.
Assume that $C_i$ ($i=1,2$) is an open curve in the null plane of $x^{-},
x^1, x^2$ and at constant $x^{+}\defineas x^{+}_i$.
The correlation function $\langle W(C_1)^\dagger W(C_2)\rangle$ can be 
calculated from the string amplitude to propagate from 
the string state corresponding to $C_1$
at $x^{+}=x^{+}_1$
to the string state corresponding to $C_2$ at $x^{+}=x^{+}_2$.
The string state corresponding to $C_1$ is such that the worldsheet fields
$(x_\perp(\sigma), \wxmin(\sigma))$ trace out $C_1$ as $0\le\sigma\le p^{+}$
and $|Y|^2\rightarrow\infty$. The state is also required to have
the specified R-charge corresponding to its dipole vector opening.
Note that in this prescription we use the $AdS_5\times S^5$ Hamiltonian
${\cal H}$ given in equation (\ref{MTlcH}) and not the deformed one
(${\cal H} +{\cal H}_1$).

%
%
%
%
%
%
%
%

\subsection{Open Wilson surfaces in generalized twisted theories}
Similarly, we can discuss correlation functions of open Wilson
surfaces in discpole theories.
Since the total R-charge of an expression
$\langle W(S_1)\cdots W(S_r)\rangle$ must be zero, the sum of 
the areas of the openings in the surfaces must be zero.
(Note that the surfaces are oriented and the areas are therefore signed.)
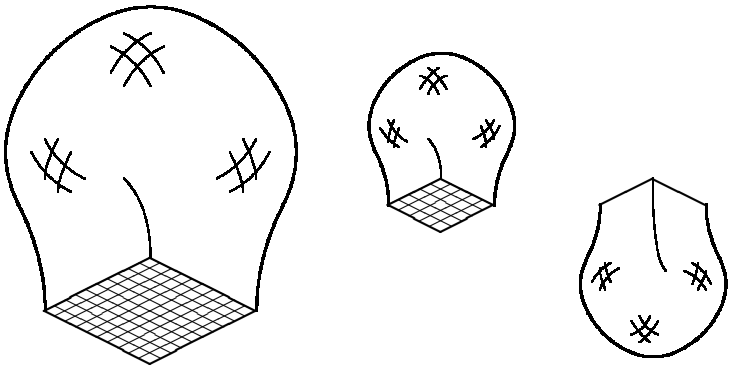
\begin{figure}[ht]
\begin{picture}(400,140)

\thicklines

\put(40,10){\begin{picture}(110,140)
  \thicklines
  \put(40,0){\line(-2,1){40}} %
  \put(0,20){\line(2,1){40}} %
  \put(40,40){\line(2,-1){40}} %
  \put(80,20){\line(-2,-1){40}} %
  \qbezier(80,20)(80,40)(90,60) 
  \qbezier(0,20)(0,40)(-10,60) 
  \qbezier(-10,60)(-20,80)(-10,100) 
  \qbezier(90,60)(100,80)(90,100) 
  \qbezier(90,100)(80,120)(60,130) 
  \qbezier(-10,100)(0,120)(20,130) 
  \qbezier(60,130)(40,140)(20,130) 

  \thinlines
  \put(35,105){\begin{picture}(40,40)
    \qbezier(5,20)(-5,15)(-10,5) 
    \qbezier(10,15)(0,10)(-5,0) 
    \qbezier(-5,20)(5,15)(10,5) 
    \qbezier(-10,15)(0,10)(5,0) 
    \end{picture}}

  \put(5,65){\begin{picture}(40,40)
    \qbezier(0,20)(-5,10)(-5,5) 
    \qbezier(5,15)(0,5)(0,0) 
    \qbezier(-5,20)(0,10)(10,5) 
    \qbezier(-10,15)(-5,5)(5,0) 
    \end{picture}}

  \put(85,60){\begin{picture}(40,40)
    \qbezier(-15,20)(-10,10)(-10,5) 
    \qbezier(-10,25)(-5,15)(-5,10) 
    \qbezier(0,20)(-5,10)(-15,5) 
    \qbezier(-5,25)(-10,15)(-20,10) 
    \end{picture}}
  \qbezier(40,40)(40,60)(30,70) %
  \thinlines
  \multiput(40,40)(4,-2){10}{\line(-2,-1){40}} 
  \multiput(40,40)(-4,-2){10}{\line(2,-1){40}} 
  \end{picture}}

\put(170,60){\begin{picture}(55,70)
  \thicklines
  \put(20,0){\line(-2,1){20}} %
  \put(0,10){\line(2,1){20}} %
  \put(20,20){\line(2,-1){20}} %
  \put(40,10){\line(-2,-1){20}} %
  \qbezier(40,10)(40,20)(45,30) 
  \qbezier(0,10)(0,20)(-5,30) 
  \qbezier(-5,30)(-10,40)(-5,50) 
  \qbezier(45,30)(50,40)(45,50) 
  \qbezier(45,50)(40,60)(30,65) 
  \qbezier(-5,50)(0,60)(10,65) 
  \qbezier(30,65)(20,70)(10,65) 

  \thinlines
  \put(17,52){\begin{picture}(20,20)
    \qbezier(2,10)(-2,7)(-5,2) 
    \qbezier(5,7)(0,5)(-2,0) 
    \qbezier(-2,10)(2,7)(5,2) 
    \qbezier(-5,7)(0,5)(2,0) 
    \end{picture}}

  \put(2,32){\begin{picture}(40,40)
    \qbezier(0,10)(-2,5)(-2,2) 
    \qbezier(2,7)(0,2)(0,0) 
    \qbezier(-2,10)(0,5)(5,2) 
    \qbezier(-5,7)(-2,2)(2,0) 
    \end{picture}}

  \put(42,30){\begin{picture}(40,40)
    \qbezier(-7,10)(-5,5)(-5,2) 
    \qbezier(-5,12)(-2,7)(-2,5) 
    \qbezier(0,10)(-2,5)(-7,2) 
    \qbezier(-2,12)(-5,7)(-10,5) 
    \end{picture}}
  \qbezier(20,20)(20,30)(15,35) %
  \thinlines
  \multiput(20,20)(4,-2){5}{\line(-2,-1){20}} 
  \multiput(20,20)(-4,-2){5}{\line(2,-1){20}} 
  \end{picture}}

\put(250,80){\begin{picture}(55,70)
  \thicklines
  \put(20,0){\line(-2,-1){20}} %
  \put(40,-10){\line(-2,1){20}} %
  \qbezier(40,-10)(40,-20)(45,-30) 
  \qbezier(0,-10)(0,-20)(-5,-30) 
  \qbezier(-5,-30)(-10,-40)(-5,-50) 
  \qbezier(45,-30)(50,-40)(45,-50) 
  \qbezier(45,-50)(40,-60)(30,-65) 
  \qbezier(-5,-50)(0,-60)(10,-65) 
  \qbezier(30,-65)(20,-70)(10,-65) 

  \thinlines
  \put(17,-52){\begin{picture}(20,20)
    \qbezier(2,-10)(-2,-7)(-5,-2) 
    \qbezier(5,-7)(0,-5)(-2,0) 
    \qbezier(-2,-10)(2,-7)(5,-2) 
    \qbezier(-5,-7)(0,-5)(2,-0) 
    \end{picture}}

  \put(2,-32){\begin{picture}(40,40)
    \qbezier(0,-10)(-2,-5)(-2,-2) 
    \qbezier(2,-7)(0,-2)(0,0) 
    \qbezier(-2,-10)(0,-5)(5,-2) 
    \qbezier(-5,-7)(-2,-2)(2,0) 
    \end{picture}}

  \put(42,-30){\begin{picture}(40,40)
    \qbezier(-7,-10)(-5,-5)(-5,-2) 
    \qbezier(-5,-12)(-2,-7)(-2,-5) 
    \qbezier(0,-10)(-2,-5)(-7,-2) 
    \qbezier(-2,-12)(-5,-7)(-10,-5) 
    \end{picture}}
  \qbezier(20,0)(20,-30)(25,-35) %
  \thinlines
  \end{picture}}

\end{picture}
\caption{\it Correlation function of Open Wilson surfaces.
For a nonzero value the orientation of the surfaces must be
such that the total oriented area of all the openings is zero.}
\end{figure}

In more general twisted theories we expect to find operators
that are parameterized by manifolds with a boundary.
For example, in the theories that are obtained by probing
the background 
$$
\VD{1\dots p}-2\pi\hM_{ij}\QJ{ij}\in\Z,
$$
given by equations (\ref{TSwmet})-(\ref{TSBphi}) (with $p\le 4$),
we expect to find operators that are parameterized by
$(p+1)$-dimensional manifolds with $p$-dimensional boundaries
that are restricted to be in a $p$-dimensional hyperplane
that is parallel to $x_1,\dots,x_p$.

\subsection{Open Wilson surfaces from the SUGRA dual}
In subsection (\ref{subsec:OWLD}) we explained the nonlocality
of the dipole theory using T-dual variables.
However, an attempt to explain the nonlocality of the discpole
theories along similar lines fails as we shall now see.
Referring to equation (\ref{M5D}),
we can write $S^4$ as a $T^2$ fibration over a base
$S^2$ (with a ring of singular fibers),
$$
\nv = (
\yb_3\,
\yb_1\sin\fb_1,\,
\yb_1\cos\fb_1,\,
\yb_2\sin\fb_2,\,
\yb_2\cos\fb_2),
$$
with $\yb_1^2 + \yb_2^2 +\yb_3^2 = 1,$ where
$(\yb_1, \yb_2, \yb_3)$ parameterize the base $S^2$.
We will take a generic
\be\label{hMDPgen}
\hM=
\left(\begin{array}{llllll}
0 & 0 & 0 & 0 & 0 \\
0 & 0 & L_1 & 0 & 0 \\
0 & -L_1 & 0 & 0 & 0 \\
0 & 0 & 0 & 0 & L_2 \\
0 & 0 & 0 & -L_2 & 0 \\
\end{array}\right)
\ee
The supersymmetric case (\ref{M5D}) is $L_1 = L_2 = \bar{L}$.
The metric and $C$-field (\ref{M5D}) can now be rewritten as
\bear
 l_p^{-2} ds^{2} &=& \frac{u}{R^{\frac{1}{2}}}
     \left(dx_{+}dx_{-} -dx_{1}^{2} - dx_{2}^{2} - dx_{3}^{2} 
-\frac{dx_4^2 + dx_5^2}{1+u^2\sum_{i=1}^2 L_i^2 \yb_i^2}\right)
-R^2 \left(d\Omega_2^2 +\frac{du^2}{u^2}\right)
\nn\\ &&
-R^2 \left\lbrack\sum_{i=1}^2 \yb_i^2 d\fb_i^2 
-\frac{u^2}{1+u^2\sum_{i=1}^2 L_i^2 \yb_i^2}
\left(\sum_{i=1}^2 L_i \yb_i^2 d\fb_i\right)^2\right\rbrack\;,
\nn\\
C &=& 
\frac{u^2}{1+u^2\sum_{i=1}^2 L_i^2 \yb_i^2}
  \sum_{i=1}^2 L_i \yb_i^2 d\fb_i\wdg dx_4\wdg dx_5,\nn
\label{dsGDiscP}
\eear
where we defined $R\defineas (\pi N)^{\frac{1}{3}}$.
Let us make a few observations about this metric.
Set 
$$
\hDen(u,y)^2\defineas 1 + u^2 (L_1^2\yb_1^2 +L_2^2\yb_2^2).
$$
For fixed $\yb_1, \yb_2, \yb_3$ the coordinates $\fb_1, \fb_2$
describe a $T^2$ with metric given by
\bear
ds^2 &=& \frac{R^2}{\hDen(u,y)^2}\left\lbrack
(1+u^2 L_2^2\yb_2^2)\yb_1^2 d\fb_1^2
+(1+u^2 L_1^2\yb_1^2)\yb_2^2 d\fb_2^2
+ 2 u^2 L_1 L_2 \yb_1^2\yb_2^2 d\fb_1 d\fb_2
\right\rbrack.
\nn\\ &&
\eear
This $T^2$ has area and complex structure:
$$
\sqrt{\det G} = \left|\frac{R^2 \yb_1 \yb_2}{\hDen(u,y)}\right| l_p^2,
\qquad
\tau = \frac{1}{1+u^2 L_2^2 \yb_2^2}\left(
u^2 L_1 L_2 \yb_2^2 + i\frac{\yb_2}{\yb_1}\hDen(u,y)
\right).
$$
We are interested in the behavior of the metric near the boundary
$u\rightarrow\infty$.
In that limit
for generic $\yb_1, \yb_2, \yb_3$ (i.e. away from the singular ring)
we have $\hDen\rightarrow\infty$. Therefore the area of the $T^2$ shrinks
to zero on the boundary and the complex structure $\tau$ becomes real.
A duality to type-IIB is not helpful because because real $\tau$ is
a singular limit of type-IIB. The origin of nonlocality therefore remains
mysterious!

\section{Discussion}\label{sec:discussion}
We have argued that many field theories that are realized on branes
have a rich class of deformations that break Lorentz invariance and
locality. Super-Yang-Mills theory has a ``dipole deformation'' for which
the nonlocality is parameterized by vectors, the $(2,0)$ theory
has a ``discpole deformation'' for which the nonlocality is parameterized
by 2-forms and little string theory has various deformations parameterized
by vectors, 3-forms, or 5-forms in type-IIB 
[corresponding to the cases $p=4,2,0$ in (\ref{SUGNS5}) respectively]
and by 2-forms and 4-forms in type-IIA
[corresponding to the cases $p=3,1$ in (\ref{SUGNS5}) respectively].
We have studied the large $N$ limit supergravity duals of these deformed
field theories and we have seen that they simplify when the deformation
parameters are lightlike.
We have analyzed Wilson loops and surfaces in the corresponding theories
and we calculated the quark anti-quark potentials for the gauge theories
and string anti-string tensions for the $(2,0)$ theory.
The contribution of the (vector or tensor)
deformation to the potential can be clearly
seen as a subleading term in (\ref{DIDI}),(\ref{QQAllCases}).
In the case of a lightlike deformation
we were able to isolate the signature of the deformation
as the leading contribution
to the potential of a certain
(quark or stringlike) object anti-object configuration
[see equations (\ref{QQELL}),(\ref{QQELLDP})].

The deformed theories also have open Wilson lines and (presumably)
open Wilson surfaces.
We calculated simple correlation functions of the open Wilson lines.
Here again there is a simplification for the lightlike deformations.
We showed that the worldsheet lightcone
Hamiltonian of a string in the supergravity
dual of the deformed (dipole) theory is essentially identical to 
the worldsheet lightcone Hamiltonian for (undeformed) $AdS_5\times S^5$
[see equation (\ref{MTlcH})]. The only difference is in the boundary conditions
on the fields [see equations (\ref{Newbc}),(\ref{wxminbc})].
These modified boundary conditions also explain the nonlocality in the
theory and the opening in the Wilson lines.
Understanding the nonlocality of the tensor deformations of the $(2,0)$
and little string theories still remains a challenge!

\acknowledgments
It is a pleasure to thank Petr Horava, Andrei Mikhailov,
Mukund Rangamani, Radu Tatar, Uday Varadarajan and Hossein Yavartanoo
for helpful discussions.
The work of O.J.G was supported by the
Berkeley center for Theoretical Physics, by the NSF grant PHY-0098840, and
by DOE Grant DE-AC03-76SF00098.

\appendix
\section{A brief review of dipole theories}\label{app:review}
For the sake of being self-contained we will present a brief review
of dipole theories. We refer the reader to
\cite{Bergman:2001rw,Alishahiha:2002ex,Ganor:2002ju}
for more details.

The dipole theories are nonlocal gauge field theories that are also
not Lorentz invariant.
The gauge group can be either $SU(N)$ or $U(N)$. 
For the $U(N)$ gauge group the field contents the same as
that of $\SUSY{4}$ Super-Yang-Mills theory but the Lagrangian
is different in the following way.

First we need a constant (R-symmetry) $so(6)$-valued space-time vector $\hM$.
Similarly to the definition of Super-Yang-Mills theory on a noncommutative
$R^4$ \cite{Fairlie:1988qd}-\cite{Hoppe:1990tc},
we modify the product of two fields $\Phi_a, \Phi_b$ 
at the spacetime point $x$ to
$$
(\Phi_a\star\Phi_b)_x\defineas
e^{i\pi\langle\hM\cdot\px{y}, {\cal Q}_b\rangle
  -i\pi\langle\hM\cdot\px{z}, {\cal Q}_a\rangle}
\left(\Phi_a(y)\Phi_b(z)\right)|_{y=z=x}
$$
Here ${\cal Q}$ is the operator R-symmetry charge which takes values in
$so(6)$. $\langle \cdot,\cdot\rangle$ is the Killing form on $so(6)$ and
the $so(6)$-valued product $\hM\cdot\partial$
denotes the scalar product of $\hM$ and $\partial$
as spacetime vectors.
The $\star$-product is associative if all $4$ $so(6)$-valued components
of the spacetime vector $\hM$ commute.
We now replace all products with $\star$-products.

Note that if $\Phi$ is a field such that
$\langle\hM,{\cal Q}\rangle\Phi = i L\Phi$, with $L$ a constant 
spacetime vector, we say that $\Phi$ has {\it dipole-vector} $2\pi L$.
This is justified by the expression for the covariant derivative
$$
(D_\u\Phi)_x \defineas
\px{\u}\Phi(x) -i (A_\u\star\Phi)_x +i (\Phi\star A_\u)_x
=
\px{\u}\Phi(x) -i A_\u(x-\pi L)\Phi(x)
+i \Phi(x) A_\u(x+\pi L).
$$
{}From the Lagrangian of the $U(N)$ dipole theory one can obtain
the $SU(N)$ dipole theory by freezing the gauge field corresponding
to the $U(1)$ center of the gauge group. Note, however, that we
cannot impose the tracelessness condition on the scalar
and spinor field combinations with nonzero dipole vectors.
For example, a field $\Phi(x)$ with dipole vector $2\pi\vL$
does not transform in the adjoint representation
of the gauge group but rather in the $(N,\overline{N})$ representation
of the product group $U(N)_{x-\pi\vL}\times U(N)_{x+\pi\vL}$
where $U(N)_x$ is the gauge group at the spacetime point $x$.

\end{document}